\documentclass[twocolumn,prl,superscriptaddress,floats,nobibnotes,notitlepage,citeautoscript]{revtex4-1}

\usepackage[pdftex]{graphicx,hyperref}
\usepackage[utf8]{inputenc}
\usepackage{amsmath,bm}
\usepackage{bbold}
\usepackage{color,soul}
\usepackage{xstring}
\usepackage{wasysym}
\usepackage{xspace}
\usepackage{amssymb}
\usepackage{time}
\usepackage{bbold}
\usepackage{subfigure}
\usepackage{multirow}
\usepackage{xspace}
\usepackage{url}

\setcitestyle{super}

\def\<{\langle}
\def\>{\rangle}
\DeclareMathOperator{\Tr}{Tr}

\newcommand{\eg}[0]{e.g.\@\xspace}

\newcommand{\ve}[1]{\boldsymbol{#1}}

\begin{document}

\title{Superconductivity from  the Condensation of  Topological  Defects\\ in a Quantum Spin-Hall Insulator}

\author{\firstname{Yuhai} \surname{Liu}}
\affiliation{\mbox{Department of Physics, Beijing Normal University, Beijing 100875, China }}
\author{\firstname{Zhenjiu} \surname{Wang}}
\affiliation{\mbox{Institut f\"ur Theoretische Physik und Astrophysik, Universit\"at W\"urzburg, Am Hubland, 97074 W\"urzburg, Germany}}
\author{\firstname{Toshihiro} \surname{Sato}}
\affiliation{\mbox{Institut f\"ur Theoretische Physik und Astrophysik, Universit\"at W\"urzburg, Am Hubland, 97074 W\"urzburg, Germany}}
\author{\firstname{Martin} \surname{Hohenadler}}
\affiliation{\mbox{Institut f\"ur Theoretische Physik und Astrophysik, Universit\"at W\"urzburg, Am Hubland, 97074 W\"urzburg, Germany}}
\author{\firstname{Chong} \surname{Wang}}
\affiliation{\mbox{Perimeter Institute for Theoretical Physics,
Waterloo, Ontario, Canada N2L 2Y5}}
\author{\firstname{Wenan} \surname{Guo}}
\email{waguo@bnu.edu.cn}
\affiliation{\mbox{Department of Physics, Beijing Normal University, Beijing 100875, China }}
\affiliation{\mbox{Beijing Computational Science Research Center, Beijing 100193, China}}
\author{\firstname{Fakher F.} \surname{Assaad}}
\email{fakher.assaad@physik.uni-wuerzburg.de}
\affiliation{\mbox{Institut f\"ur Theoretische Physik und Astrophysik, Universit\"at W\"urzburg, Am Hubland, 97074 W\"urzburg, Germany}}

\begin{abstract}
  The discovery that spin-orbit coupling can generate a new state of matter
  in the form of quantum spin-Hall (QSH) insulators has brought topology to
  the forefront of condensed matter physics. While QSH states from spin-orbit
  coupling can be fully understood in terms of band theory, fascinating
  many-body effects are expected if the state instead results from
  interaction-generated symmetry breaking. In particular, topological defects
  of the corresponding order parameter provide a route to exotic quantum
  phase transitions. Here, we introduce a model in which the condensation of
  skyrmion defects in an interaction-generated QSH insulator produces a
  superconducting (SC) phase. Because vortex excitations of the latter carry
  a spin-$1/2$ degree of freedom numbers, the SC order may be understood as
  emerging from a gapless spin liquid normal state. The QSH-SC transition is
  an example of a deconfined quantum critical point (DQCP), for which we
  provide an improved model with only a single length scale that is
  accessible to large-scale quantum Monte Carlo simulations.
\end{abstract}

\maketitle

\begin{figure}[b]
  \includegraphics[width=0.45\textwidth]{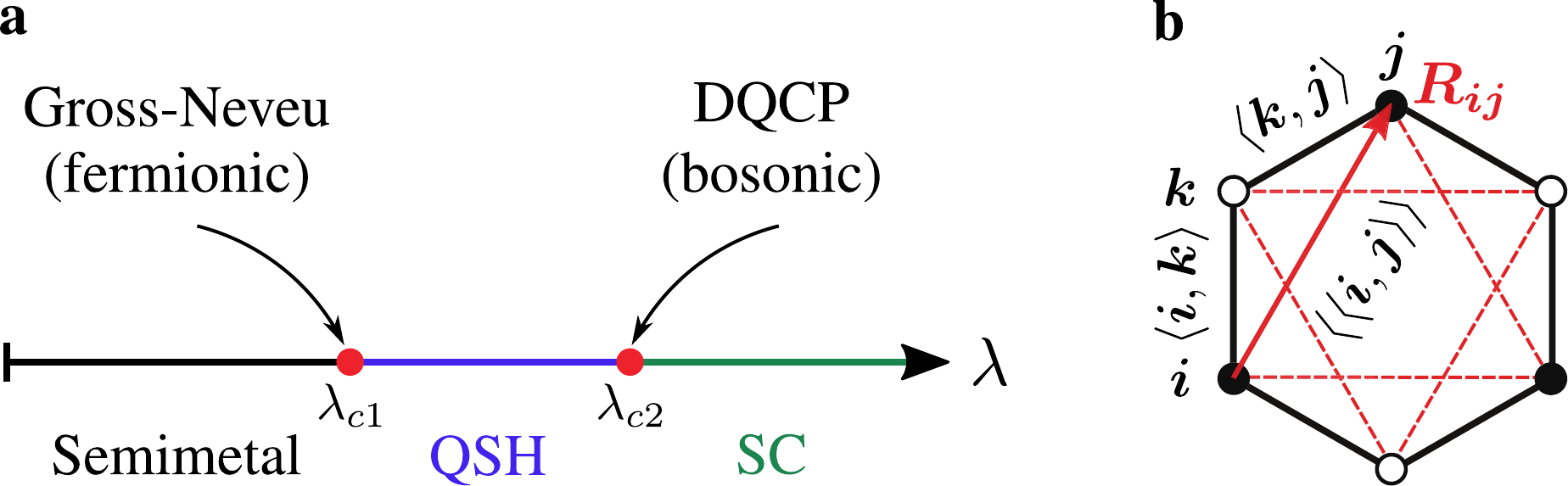}
  \caption{\label{fig:modelandphasediagram}
    {\bf Phase diagram and model.} {\bf a} Schematic ground-state phase
    diagram with semimetallic, QSH, and SC phases.  {\bf b} Illustration of
    nearest- and next-nearest neighbours and the vector $\boldsymbol{R}_{ij}$ on a
    plaquette of the honeycomb lattice.
 }
\end{figure}

In the Kane-Mele model for the QSH insulator,\cite{KaneMele05b} the original
SU(2) spin symmetry is explicitly broken by spin-orbit coupling. Here, we
instead consider the case where this symmetry is preserved by the Hamiltonian
but spontaneously broken by an interaction-generated QSH state.\cite{Raghu08}
At the mean-field level, the latter is characterised by an SO(3) order
parameter constant in space and time and a band structure with a non-trivial
$\mathbb{Z}_2$ topological index \cite{KaneMele05b,Koenig07,Reis17}.
Long-wavelength fluctuations of this order parameter include in particular
the Goldstone modes that play a key role for phase transitions to, \eg, a
Dirac semimetal. Such a transition, illustrated in
Fig.~\ref{fig:modelandphasediagram}a, is described by a Gross-Neveu-Yukawa
field theory \cite{Gross74} with QSH order encoded in a mass in the
underlying Dirac equation.  Fluctuations can also take the form of
topological (`skyrmion') defects that correspond to a nontrivial winding of
the order parameter vector. Due to the topological nature of the QSH state
itself, such skyrmions carry electric charge 2e:\cite{Grover08} as
illustrated in the supplemental material, the insertion of a skyrmion in
a system with open boundaries pumps a pair of charges from the valence to the
conduction band through the helical edge states. The condensation of skyrmion
defects---which coincides
with the destruction of the QSH state---represents a new route to generate a
SC state. In contrast to a weakly coupled Bardeen-Cooper-Schrieffer-type SC,
its vortices enclose a spin-$1/2$ degree of freedom corresponding to the
fractionalized QSH order parameter \cite{Grover08}.

A direct QSH-SC phase transition (Fig.~\ref{fig:modelandphasediagram}a)
is an instance of a DQCP \cite{Senthil04_1,Senthil04_2,WangC17},
the concept of which relies on the topological defects of one phase carrying
the {\it charge} of the other phase. Defect condensation then
provides a mechanism for a continuous transition between two states with
different broken symmetries (SO(3) for QSH, U(1) for SC) that is forbidden by
Landau theory. Despite considerable numerical efforts \cite{Shao15,Nahum15},
DQCPs remain a subject of intense debate.   Important questions
include their  very nature---weakly first order or continuous \cite{WangC17}---and the role
of emergent symmetries \cite{Nahum15_1}.  One of the difficulties lies in the
fact that previous lattice realizations\cite{Kawashima07,Sandvik07,Nahum15} involve
antiferromagnetic (AFM)  and valence bond solid  (VBS)  phases. For the
widely studied square lattice, the VBS state breaks the discrete $\mathbb{Z}_4$
rotation symmetry, whereas the field theory has a U(1) symmetry. The latter
is recovered on the lattice exactly at the critical point, but in general
the $\mathbb{Z}_4$ symmetry breaking term is relevant. The additional
length scale at which the $\mathbb{Z}_4$ symmetry becomes visible obscures the
numerical analysis. In the field theory, this translates  into the notion that
quadruple  skyrmion addition (monopole)
events of the AFM SO(3) order parameter are  irrelevant at criticality but
proliferate slightly away from this point to generate the VBS state
\cite{Senthil04_1}. Hence, the theory is subject to a dangerously irrelevant
operator. This complication is completely avoided in the model introduced
here, where the DQCP separates QSH and SC rather than AFM and VBS phases.
QSH and AFM order are both described by an SO(3) order parameter. However,
instead of the $\mathbb{Z}_4$ symmetry broken by lattice VBS state, the SC phase
breaks the same global U(1) gauge symmetry (charge conservation) on the lattice and in the continuum.
Therefore, the number of skyrmion defects with charge 2e is conserved and
monopoles are absent.

The exciting prospects of (i) SC order from topological defects of a spontaneously
generated QSH state and (ii) a monopole-free realisation of a DQCP motivate the
search for a suitable lattice model amenable to quantum Monte Carlo
simulations without a sign problem. Such efforts are part of the recent surge
of designer Hamiltonians aimed at studying exotic phases and phase transitions \cite{Berg12,Kaul13,Xu16c,Assaad16,Gazit16,Gazit18,SatoT17}.  Here, we
start from a tight-binding model of Dirac fermions in the form of electrons
on the honeycomb lattice with nearest-neighbour hopping (see
Fig.~\ref{fig:modelandphasediagram}b), as described by
\begin{equation}\label{eq:hamiltonian1}
\hat{H}_t=-t\sum_{\langle \ve{i}, \ve{j} \rangle}(\hat{\boldsymbol{c}}^{\dag}_{\ve{i}} \hat{\boldsymbol{c}}^{}_{\ve{j}}+\text{H.c.}).
\end{equation}
The spinor $\hat{\boldsymbol{c}}^{\dag}_{\ve{i}} =
\big(\hat{c}^{\dag}_{\ve{i},\uparrow},\hat{c}^{\dag}_{\ve{i},\downarrow}
\big)$, where $\hat{c}^{\dag}_{\ve{i},\sigma} $ creates an electron at
lattice site $\ve{i}$ with spin $\sigma$. Equation~(\ref{eq:hamiltonian1})
yields the familiar graphene band structure with gapless, linear excitations
at the Dirac points.\cite{Novoselov05} A suitable interaction that generates
the above physics is
\begin{equation}\label{eq:hamiltonian2}
\hat{H}_{\lambda}=-\lambda\sum_{\hexagon} \Bigg( \sum_{ \langle \langle \ve{i},\ve{j} \rangle \rangle  \in \hexagon } \mathrm{i}  \nu_{\boldsymbol{ij}}
 \hat{\boldsymbol{c}}^{\dagger}_{\ve{i}} \boldsymbol{\sigma} \hat{\boldsymbol{c}}^{}_{\ve{j}}+\text{H.c.} \Bigg)^2.
\end{equation}
The first sum is over all the hexagons of a honeycomb lattice with $L\times
L$ units cells and periodic boundary conditions. The second sum is over all
pairs of next-nearest-neighbour sites of a hexagon, see
Fig.~\ref{fig:modelandphasediagram}b. The quantity
$\nu_{\boldsymbol{ij}}=\pm1$ is identical to the Kane-Mele model\cite{KaneMele05b}; for a
path from site $\boldsymbol{i}$ to site $\boldsymbol{j}$ (connected by $\boldsymbol{R}_{\boldsymbol{ij}}$, see
Fig.~\ref{fig:modelandphasediagram}b) via site $\boldsymbol{k}$, $\nu_{\boldsymbol{ij}}={\hat{\boldsymbol{e}}_z\cdot(\boldsymbol{R}_{\boldsymbol{ik}}\times\boldsymbol{R}_{\boldsymbol{kj}})}/{|\hat{\boldsymbol{e}}_z\cdot(\boldsymbol{R}_{\boldsymbol{ik}}\times\boldsymbol{R}_{\boldsymbol{kj}})|}
$ with $\hat{\boldsymbol{e}}_z$ a unit vector perpendicular to the honeycomb plane.
Finally, $ \boldsymbol{\sigma}=(\sigma^x,\sigma^y,\sigma^z)$ with the Pauli spin
matrices $\sigma^{\alpha}$.  The rationale for this choice of interaction is easy
to understand.  Without the square, and taking just one of the  three
Pauli matrices, equation~(\ref{eq:hamiltonian2}) reduces to the Kane-Mele
spin-orbit coupling that explicitly breaks SO(3) spin symmetry. In contrast,
this symmetry is preserved by $\hat{H}_\lambda$ but spontaneously broken by
long-range QSH order. For $\lambda>0$,
the model defined by $\hat{H}=\hat{H}_t+\hat{H}_\lambda$ can be simulated
without a sign problem by auxiliary-field quantum Monte Carlo methods
\cite{Blankenbecler81,White89,Assaad08_rev}. In the following, we set
$t=1$ and consider a half-filled band with one electron per site.

\begin{figure}[t]
  \includegraphics[width=0.45\textwidth]{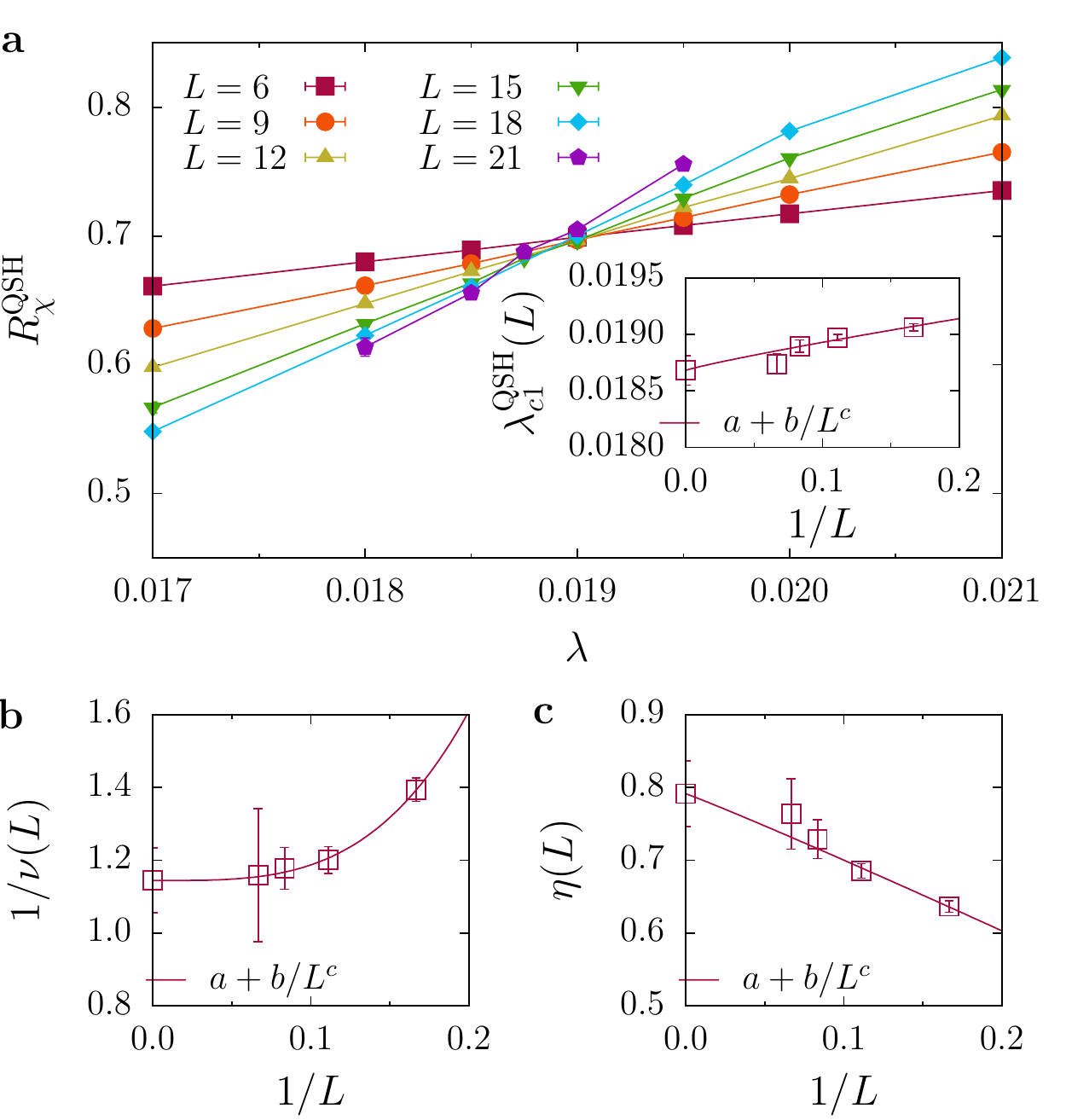}
  \caption{\label{fig:qshsm}
    {\bf Gross-Neveu semimetal-QSH transition.}
    {\bf a} Correlation ratio $R^\text{QSH}_\chi$
    [equation~(\ref{eq:correlationratio})] for different system sizes
    $L$. The extrapolation of the crossing points of $R^\text{QSH}_\chi$ for $L$
    and $L+6$ in the inset gives the critical value
    $\lambda_{c1}=0.0187(2)$. {\bf b} Finite-size scaling based on
    equation~(\ref{eq:nufromcorrelationratio}) gives an inverse
    correlation length exponent $1/\nu=1.14(9)$. {\bf c} Estimation of the
  anomalous dimension  $\eta=0.79(5)$.}
\end{figure}

A mean-field decomposition of equation~(\ref{eq:hamiltonian2}) suggests a
transition from the Dirac semimetal to a QSH state at a critical value
$\lambda_{c1}>0$. However, it is highly non-trivial if the associated
saddle point is stable. In fact, s-wave pair hopping processes
arise upon expanding the square in equation~(\ref{eq:hamiltonian2}) and
can lead to superconductivity \cite{Capponi07}. The exact phase diagram can be obtained by
quantum Monte Carlo simulations. Remarkably, as illustrated in
Fig.~\ref{fig:modelandphasediagram}a, we find two distinct
phase transitions. First, from the semimetal to a QSH state at
$\lambda_{c1}$, then from the QSH state to an s-wave SC at
$\lambda_{c2}>\lambda_{c1}$.

The semimetal-QSH transition involves the breaking of spin rotation symmetry and is
expected to be in the O(3) Gross-Neveu universality class for $N=8$ Dirac
fermions (two sublattices, two Dirac points, $\sigma=\uparrow,\downarrow$).
The local vector order parameter takes the form of a spin current,
\begin{equation}
\hat{\boldsymbol{O}}^{\text{QSH}}_{\boldsymbol{r},\boldsymbol{\delta}}  =
\mathrm{i}\hat{\boldsymbol{c}}^{\dagger}_{\boldsymbol{r}}  \boldsymbol{\sigma}
\hat{\boldsymbol{c}}^{}_{\boldsymbol{r}+ \boldsymbol{\delta}}   +  \text{H.c.},
\end{equation}
where $\ve{r}$ corresponds to a unit cell labelling a hexagon, and $\ve{r} +
\ve{\delta}$ runs over all next-nearest neighbours. Because this order
parameter is a lattice regularisation of the three QSH mass terms in the
Dirac equation, long-range order implies a mass gap.\cite{KaneMele05b} To
study the phase transition, we computed the associated susceptibility
\begin{equation}
 \chi^{O}_{\ve{\delta},\ve{\delta}'} (\boldsymbol{q})
	= \frac{1}{L^2} \sum_{\boldsymbol{r},\boldsymbol{r'}}   \int_{0}^{\beta} \text{d} \tau   e^{\mathrm{i}\boldsymbol{q}\cdot(\boldsymbol{r}-\boldsymbol{r}')} \langle  \hat{\boldsymbol{O}}_{\boldsymbol{r},\boldsymbol{\delta}}(\tau)  \hat{\boldsymbol{O}}_{\boldsymbol{r'},\boldsymbol{\delta}'}(0) \rangle.
\end{equation}
Here, we omitted the vanishing background term and concentrate on the largest
eigenvalue of  $\chi^{O}_{\ve{\delta},\ve{\delta}' } (\boldsymbol{q}) $ (see
supplementary material), henceforth denoted as $ \chi^{O}(\boldsymbol{q})$.  To detect the transition, we
consider the renormalization-group  invariant correlation ratio
\begin{equation}\label{eq:correlationratio}
1 -  \frac{\chi^{O}(\boldsymbol{Q} + \Delta \boldsymbol{q} )}{\chi^{O}(\boldsymbol{Q})} = R^{O}_\chi\left( L^{1/\nu} \left( \lambda - \lambda^O_c \right), L^{-w}  \right)
\end{equation}
with $|\Delta \boldsymbol{q}| = \frac{4\pi}{\sqrt{3} L}$,
the ordering wavevector $\boldsymbol{Q}=0$, the
correlation length  exponent $\nu$,  and the leading corrections-to-scaling
exponent  $\omega$. We set the inverse temperature $\beta = L$ in our simulations
based on the assumption of a dynamical critical exponent $z=1$.\cite{Herbut09}
In contrast to previous analyses of Gross-Neveu criticality \cite{Toldin14,Otsuka16}
we use susceptibilities rather than equal-time correlators to
suppresses background contributions to the critical fluctuations.

The results for the semimetal-QSH transition are shown
in Fig.~\ref{fig:qshsm}. The finite-size estimate of the critical value,
$\lambda^\text{QSH}_{c1}(L) $, corresponds to the crossing point of
$R^\text{QSH}_\chi$ for $L$ and $L+6$. Extrapolation to the thermodynamic
limit (inset of Fig.~\ref{fig:qshsm}a) yields
$\lambda^\text{QSH}_{c1}=0.0187(2)$. As shown in the supplementary material,
the single-particle gap is nonzero for $\lambda>\lambda^\text{QSH}_{c1}$.
The correlation length exponent was estimated from \cite{Shao15}
\begin{equation}\label{eq:nufromcorrelationratio}
  \frac{1}{\nu^{O}  (L) } =
  \left. \frac{1}{\log{r} }
    \log \left( \frac{ \frac{d} {d\lambda} R^{O}_\chi\left( \lambda, r L \right)  }{\frac{d} {d\lambda} R^{O}_\chi\left( \lambda,  L \right)  } \right)    \right|_{\lambda = \lambda^O_c(L)}
  \end{equation}
with $r = \frac{L+6}{L}$. A similar equation
can be used to determine the exponent $\eta$ from the divergence of
the susceptibility ($\chi^O \propto L^{2 - \eta }$) at criticality  (see
supplementary material).  Aside from a polynomial interpolation of the data
as a function of $\lambda$ for each  $L$, this analysis  does not require any
further fitting and, by definition, converges to the correct exponents in the
thermodynamic limit with rate $L^{-\omega}$. While previous estimates of the
critical exponents vary,\cite{Toldin14,Otsuka16,Zerf17} the values
$1/\nu = 1.14(9)$ and $\eta=0.79(5)$ from Fig.~\ref{fig:qshsm} are consistent with
$\nu = 1.02(1)$ and $\eta =  0.76(2)$ from previous work.\cite{Otsuka16} This
suggest that the semimetal-QSH transition is in the same universality class
as the semimetal-AFM transition \cite{Assaad13,Toldin14,Otsuka16}.

\begin{figure}[t]
  \includegraphics[width=0.45\textwidth]{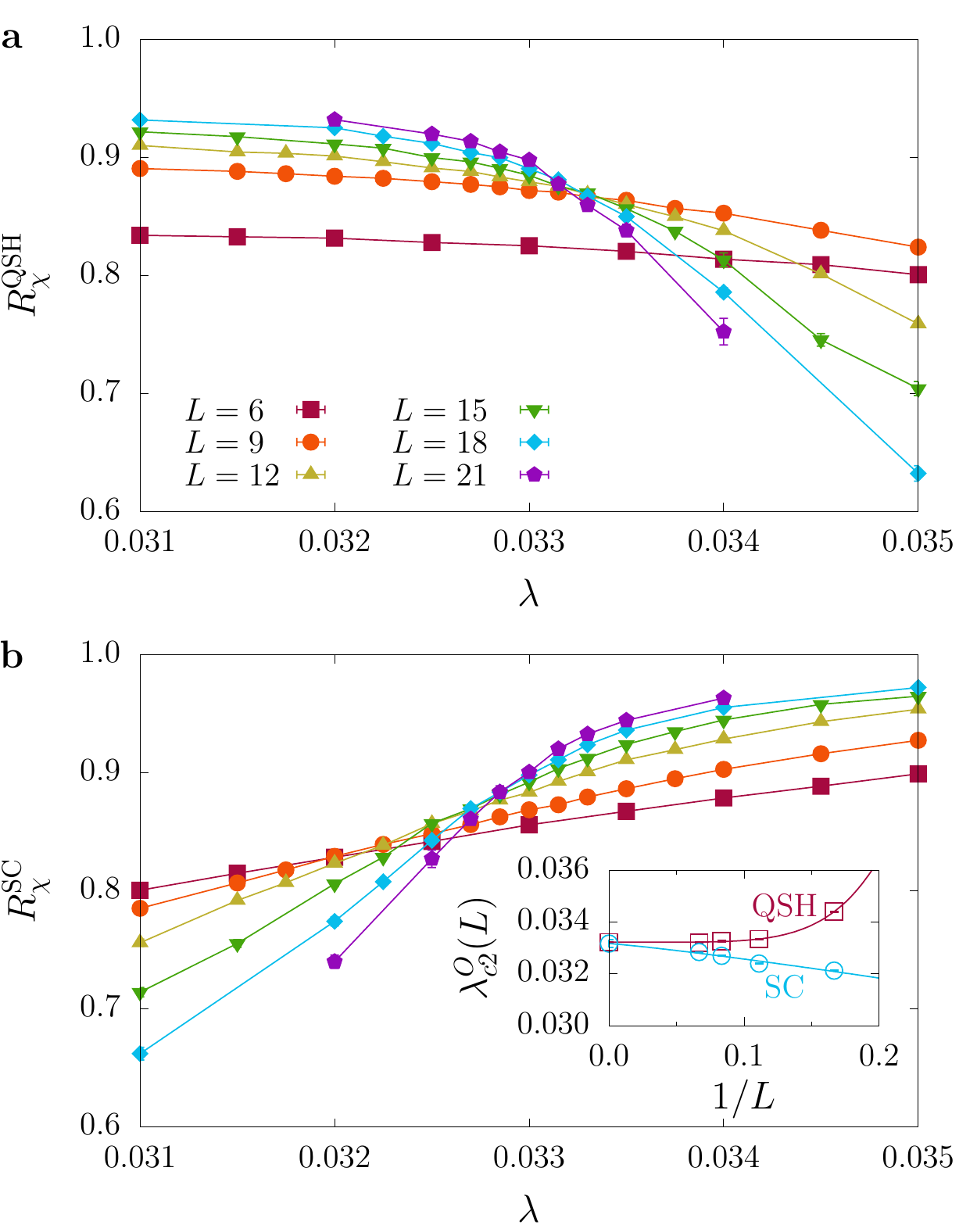}
  \caption{\label{fig:qshssc}
    {\bf Deconfined QSH-SC transition.}
    {\bf a} Correlation ratio $R^\text{QSH}_\chi$ and {\bf b} correlation
    ratio  $R^\text{SC}_\chi$ for different system sizes $L$. The
    extrapolation of the crossing points for $L$ and $L+6$  using the form
    $a+b/L^c$ (see inset of {\bf b})
    gives $\lambda_{c2}^\text{QSH}=0.03322(3)$ and
    $\lambda_{c2}^\text{SC}=0.0332(2)$.}
\end{figure}

To detect SC order,  we used the order parameter
\begin{equation}
  \hat{O}^\text{SC}_{\ve{r},\ve{\tilde{\delta}}}
  =   \frac{1}{2}
  \left( \hat{c}^{\dagger}_{\ve{r} +\ve{\tilde{\delta}},\uparrow}  \hat{c}^{\dagger}_{\ve{r} +\ve{\tilde{\delta}},\downarrow} + \text{H.c.}\right)
\end{equation}
where $\ve{r} +\ve{\tilde{\delta}}$ runs over the two orbitals of unit cell $\ve{r}$.
As before, we computed the corresponding susceptibility and used $\beta = L$ in
anticipation of $z=1$. Fig~\ref{fig:qshssc} shows that, within the very
small error bars, the critical value for SC order $\lambda_{c2}^\text{SC} =
0.0332(2)$ and the critical value for the disappearance of long-range QSH
order $\lambda_{c2}^\text{QSH} = 0.03322(3)$ are identical, suggesting a
direct QSH-SC transition. At this transition, the single-particle gap
remains of order one and we find no evidence for a first-order transition
for the available system sizes (supplementary Fig.~2).

The observed s-wave symmetry of the SC state emerges directly
from the perspective of Dirac mass terms. In $2+1$ dimensions and for $N=8$
Dirac fermions, there exist numerous quintuplets of anti-commuting mass terms
that combine different order parameters in a higher SO(5) symmetry group
\cite{Ryu09}.  A well-known example relevant for DQCPs are the
three AFM and two VBS mass terms. The three QSH mass terms form a quintuplet
with the two s-wave SC mass terms.  The resulting SO(5) order parameter allows for a
very natural derivation of the Wess-Zumino-Witten term
\cite{Abanov00,Tanaka05}, crucial for the DQCP, by integrating out the
(massive) Dirac fermions \cite{Senthil06}.

\begin{figure}[t]
  \includegraphics[width=0.45\textwidth]{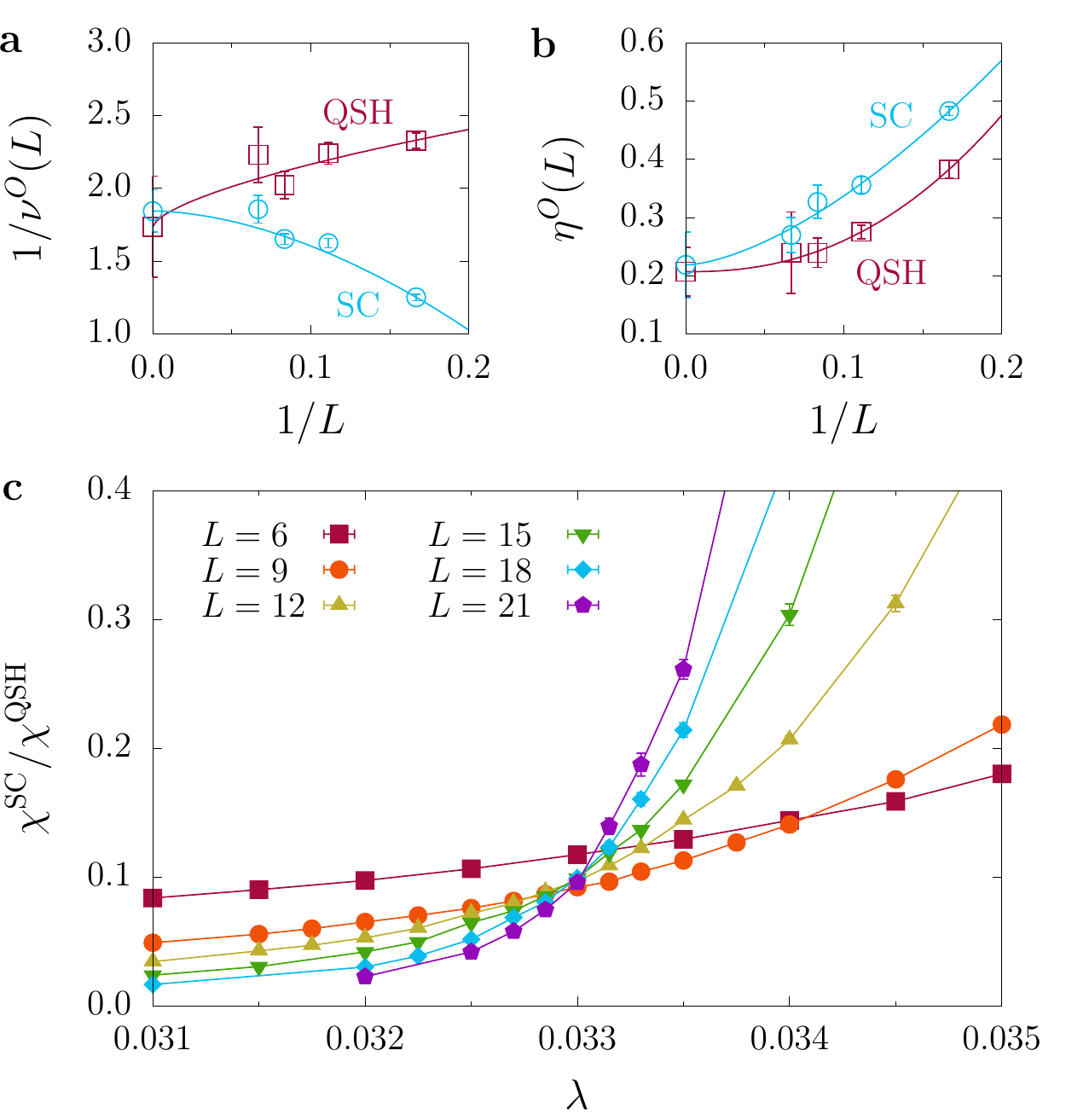}
  \caption{\label{fig:so5}
  {\bf Critical exponents for the QSH-SC transition.}
  {\bf a}, {\bf b} Critical exponents $1/\nu^\text{SC}=1.8(2)$, $1/\nu^\text{QSH}=1.7(4)$,
  $\eta^\text{SC}=0.22(6)$, and $\eta^\text{QSH}=0.21(5)$ from finite-size
  scaling of the crossing points for $L$ and $L+6$. {\bf c} Ratio of the QSH
  and SC susceptibilities for different system sizes $L$.}
\end{figure}

As argued in the introduction, the QSH-SC problem is free of monopoles, so
that our lattice model represents an improved model to study the DQCP.
Although simulations for fermions are limited to smaller system sizes than
for bosons, severe size effects due to monopoles \cite{Shao15} can be
expected to be absent. Fig~\ref{fig:so5} shows a finite-size analysis
for the correlation length exponent and the anomalous dimension, as obtained
either from the QSH or the SC correlation ratio. The resulting estimates
$\eta^{\text{QSH}} = 0.21(5)$ and $\eta^{\text{SC}} = 0.22(6)$ compare
favourably with those from loop models \cite{Nahum15} where
$\eta^{\text{AFM}} = 0.259(6)$ and $ \eta^{\text{VBS}} =0.25(3)$. An
alternative analysis in the supplementary material yields similar values.
Given the very similar anomalous dimensions $\eta^{\text{QSH}}$ and
$\eta^{\text{SC}}$ of QSH and SC fluctuations, the ratio of the QSH and SC
susceptibilities  is expected to be a renormalization group invariant, as
confirmed by Fig.~\ref{fig:so5}c. However, a crossing of different
curves at $\lambda_{c2}$ is a necessary but not sufficient condition for an
emergent SO(5) symmetry at the DQCP. In fact, a continuous transition with
emergent SO(5) symmetry can be essentially excluded here in the light of the
condition $\eta > 0.52$ from the conformal bootstrap method.\cite{Poland18}
The latter also yields a bound of $1/\nu < 1.957$   for a unitary conformal field theory with only one tuning parameter \cite{Nakayama16}
   that is satisfied by $1/\nu^{\text{SC}} =
1.8(2)$ and $1/\nu^{\text{QSH}} = 1.7(4)$ from Fig.~\ref{fig:so5}a but not
by $1/\nu = 2.24(4)$.\cite{Shao15} Simulations of the monopole-free model on
even larger lattices are required for a conclusive answer.

Our model provides a realisation of a QSH insulator emerging from
spontaneous symmetry breaking. The corresponding SO(3) order parameter
permits both long-wavelength Goldstone modes and topological skyrmion defects.
By means of a single parameter $\lambda$, we can trigger continuous quantum phase
transitions to either a semimetal or an s-wave SC state. For the
semimetal-QSH transition, the critical exponents are consistent with
Gross-Neveu universality \cite{Toldin14,Otsuka16}. The QSH-SC transition is of
particular interest since it provides a monopole-free, improved model of
deconfined quantum criticality with only one length scale. The mechanism
for SC order from the QSH state is the condensation of skyrmion defects
of the QSH order parameter with charge 2e.
For the QSH-SC transition, our values  of the anomalous dimension
match those of previous work on the AFM-VBS
transition,\cite{Sandvik07,Nahum15}  which are inconsistent with results from
conformal bootstrap studies if an SO(5) symmetry emerges at the critical
point (as is supported by numerical and analytical studies). One possible
resolution is the scenario of `pseudo-criticality' or `walking coupling
constant' \cite{Kuklov08,Nahum15,WangC17,RychkovWalking1}. In contrast, our estimate
of $1/\nu$  is  still within the conformal  bootstrap bound,\cite{Nakayama16}
although a bound-violating result is not completely ruled out given the
numerical uncertainty.  Consequently, it is of considerable
interest to exploit the full potential of quantum Monte Carlo methods in order
to access even larger lattices.  Other promising approaches that can shed
further light on DQCPs make use  of a lattice discretisation scheme based on
projection onto a Landau level that does not break continuum
symmetries.\cite{Ippoliti18}

A monopole-free realisation of DQCPs is impossible in traditional settings
because of an anomaly\cite{WangC17} associated with the SO(3)$\times$U(1)
symmetry. In the standard realisation, this anomaly is matched by the
non-onsite nature of lattice rotation symmetries\cite{MetlitskiThorngren},
but since lattice rotations are discrete, monopoles can never be completely
suppressed. Alternatively, the anomaly can be eliminated by properly
enlarging the SO(3)$\times$U(1) symmetry, essentially allowing microscopic
degrees of freedom that carry `fractional' symmetry quantum numbers. This is
what is being done in this work, where the fermions carry half-spin
and half-charge (in terms of Cooper pair charges). An even simpler
extension of the symmetry that eliminates the anomaly is SU(2)$\times$U(1),
meaning that microscopically there are charged spinless bosons, together with
both charged and neutral spin-$1/2$ bosons. A challenge for future studies is to
find a reasonably simple Hamiltonian that realises a DQCP and is amenable to
sign-free bosonic QMC simulations in, \eg, the stochastic series expansion
representation.\cite{Sandvik99b}

The SC phase generated from skyrmion defects motivates further
investigations. Its vortex excitations carry a spin-$1/2$ degree of
freedom,\cite{Grover08} so that in the quantum critical fan thermal melting
will yield a spin liquid \cite{Sandvik11}. It is also possible to add an independent
attractive Hubbard interaction to explore a semimetal-QSH-SC tricritical
point (as opposed to the recently discovered semimetal-AFM-VBS tricritical point\cite{SatoT17})
with predicted SO(5) Gross-Neveu criticality.\cite{Janssen18} The vector
form of $\hat{H}_\lambda$ makes it straight forward to reduce the
SO(3) QSH symmetry to U(1) and thereby investigate easy-plane realisation of
DQCPs with a U(1)$\times$U(1) symmetry on the lattice.  Work along these
directions is in progress.

\subsection*{Methods}

We employed the ALF\cite{ALF_v1} implementation of the
auxiliary-field finite-temperature quantum Monte Carlo
method \cite{Blankenbecler81,White89,Assaad08_rev}. The interaction term is
written as a perfect square with negative prefactor ($\lambda > 0$), allowing
for a decomposition in terms of a real Hubbard-Stratonovitch field.  For each
field configuration, time-reversal symmetry holds and the eigenvalues of the
fermion matrix occur in complex conjugate pairs \cite{Wu04,Li16,Wei16}.
At low temperatures, the scales of the imaginary time propagation do not fit into double
precision real numbers and we have used methods to circumvent this issue.\cite{Hofmann18}
The imaginary-time discretisation was $\Delta\tau=0.2$. For reasons explained in
the supplementary material, we chose a symmetric Trotter decomposition
that minimises discretization errors.

\subsection*{Acknowledgements}
We acknowledge illuminating discussions with T. Grover, and with J. Hofmann
regarding extensions of the ALF\cite{ALF_v1} package. We thank the
Gauss Centre for Supercomputing (SuperMUC at the  Leibniz Supercomputing
Centre) for generous allocation of supercomputing resources as well as
financial support from the Deutsche Forschungsgemeinschaft  (DFG) grant
AS120/14-1 for further development of the ALF package.  TS thanks   the DFG
for financial support from grant number AS120/15-1.  MH and ZW  are
supported by the DFG collaborative research centre SFB1170
ToCoTronics (project C01), YL and WG by NSFC under grants nos.~11775021 and
11734002. Research at Perimeter Institute (CW) is supported by the Government
of Canada through the Department of Innovation, Science and Economic
Development Canada and by the Province of Ontario through the Ministry of
Research, Innovation and Science.

\subsection*{Author contributions}

CW, WG and FFA  conceived the project.   YL and ZW implemented  the model in
the ALF package and carried out the simulations  under the guidance of TS and
MH.  All authors participated in the interpretation of the results and the
writing of the manuscript.

\subsection*{Competing financial interests}

The authors declare no competing financial interests.


\onecolumngrid

\setcounter{figure}{0}
\renewcommand{\figurename}{S.~Fig.}

\section{Supplemental Material}

\maketitle


\subsection{Observables}
\label{sec:Observables}

Symmetry-broken states are  characterised by a local order parameter $\hat{\ve O}_{\ve{r}, \ve{\delta}}$,
where $\ve{r}$ denotes a unit cell and $\ve{\delta}$ an orbital within the unit cell.
The associated  time-displaced correlation functions read
\begin{equation}
S^O_{\ve{\delta}, \ve{\delta'}}(\ve{q}, \tau)=\frac{1}{L^2}\sum_{ \ve{r},\ve{r'} } \langle \hat{ \ve{O}}_{\ve{r},\ve{\delta}}(\tau) \cdot \hat{\ve{O}}_ {\ve{r'}, \ve{\delta'}}(0) \rangle  e^{i \ve{q} \cdot (\ve{r}- \ve{r')} }  \;.
\end{equation}
For the finite-size scaling analysis, we consider the order parameter
\begin{equation}
   m^O  = \sqrt{\frac{{\Lambda_1} (S^O_{\ve{\delta},\ve{\delta'}} (\ve{0}, 0))}{L^2}}\,,
\end{equation}
the equal-time correlation ratios
\begin{equation}
   R^O = 1 - \frac{  {\Lambda_1} (S^O_{\ve{\delta},\ve{\delta'}} ( \Delta \ve{q}, 0)) }{  {\Lambda_1} (S^O_{\ve{\delta},\ve{\delta'}} (\ve{0}, 0))  }\,,
\end{equation}
with $|  \Delta \ve{q}|  = \frac{4 \pi}{\sqrt{3} L} $,  the susceptibilities
\begin{equation}
   \chi^O  =   \Lambda_1 \left( \mbox{$\int_{0}^{\beta}$}  d\tau  S^O_{\ve{\delta},\ve{\delta'}}(\ve{0}, \tau) \right) \,,
\end{equation}
and the corresponding correlation ratios
\begin{equation}
 R_\chi^O =  1 - \frac{   \Lambda_1 \left( \int_{0}^{\beta}  d\tau  S_{\ve{\delta},\ve{\delta'}}(\Delta  \ve{q},\tau) \right) }{\Lambda_1\left( \int_{0}^{\beta}  d\tau  S^O_{\ve{\delta},\ve{\delta}}(\ve{0},\tau) \right) }\,.
\end{equation}
Here, $\Lambda_1()$ indicates the largest eigenvalue of the corresponding
matrix in orbital space ($6 \times 6$ for spin currents, $2 \times 2$ for pairing)
and the ordering wave vector is at the $\Gamma$ point.

These quantities exhibit the following finite-size scaling behaviour near the critical point:
\begin{eqnarray}\label{FSS}\nonumber
  m^O   (L, \lambda)  &=&  L^{(2-d-z-\eta)/{2}}  f_1({L^z}/{\beta} , (\lambda - \lambda_c ) L^{1/\nu},  L^{-\omega_1} )\,, \\ \nonumber
  R^O   (L, \lambda)  &=&  f_2({L^z}/{\beta} , (\lambda - \lambda_c ) L^{1/\nu} ,  L^{-\omega_2} )\,,  \\\nonumber
 \chi^O (L, \lambda)  &=&  L^{2-{\eta}}  f_3({L^z}/{\beta} ,  (\lambda - \lambda_c ) L^{1/\nu} ,  L^{-\omega_3} )\,, \\
  R_\chi^O (L, \lambda) &=&  f_4({L^z}/{\beta} , ( \lambda - \lambda_c ) L^{1/\nu} ,  L^{-\omega_4} )\,.
\end{eqnarray}
Here, $\lambda_c, \nu, \eta$, and $z$ are the critical coupling, the
correlation length exponent, the anomalous dimension, and the dynamical
critical exponent, respectively.

The correlation ratios $R^O$ and $R_\chi^O$ are both renormalization group (RG)
invariant quantities at the critical point and hence provide a simple way
to estimate $\lambda_c$ and $\nu$ without any knowledge about
$\eta$. However, the generic corrections-to-scaling
exponent $\omega$ is not necessarily the same for all four quantities in
equation~(\ref{FSS}). Such corrections generally arise from irrelevant operators
of the fixed point and the analytic part of the free energy. If the absolute
value of the negative RG dimension is relatively large, the main
contribution to $\omega$ will come from the
background term of the free energy \cite{Toldin14}. In this case,
\begin{equation}
   \omega_1 = \omega_2 = 2 - z - \eta,  \  \  \   \omega_3 = \omega_4 = 2 - \eta\,.
\end{equation}
This suggests that the susceptibility $\chi$ and the corresponding
correlation ratio $R_\chi$ will have smaller scaling corrections than the
corresponding equal-time quantities if the effect of the negative RG
dimension is small at $\lambda_c$.

We assumed a dynamical critical exponent $z=1$ for both the SM-QSH and the
QSH-SC transition. This is motivated by the Lorentz
invariance of the corresponding field theories
\cite{Senthil04_2,Gross74}. Accordingly, in our simulations, we used $\beta=L$
and thereby fixed $L^z/\beta$ in the above finite-size scaling expressions.

\begin{figure}
\centering
\includegraphics[width=0.9\textwidth]{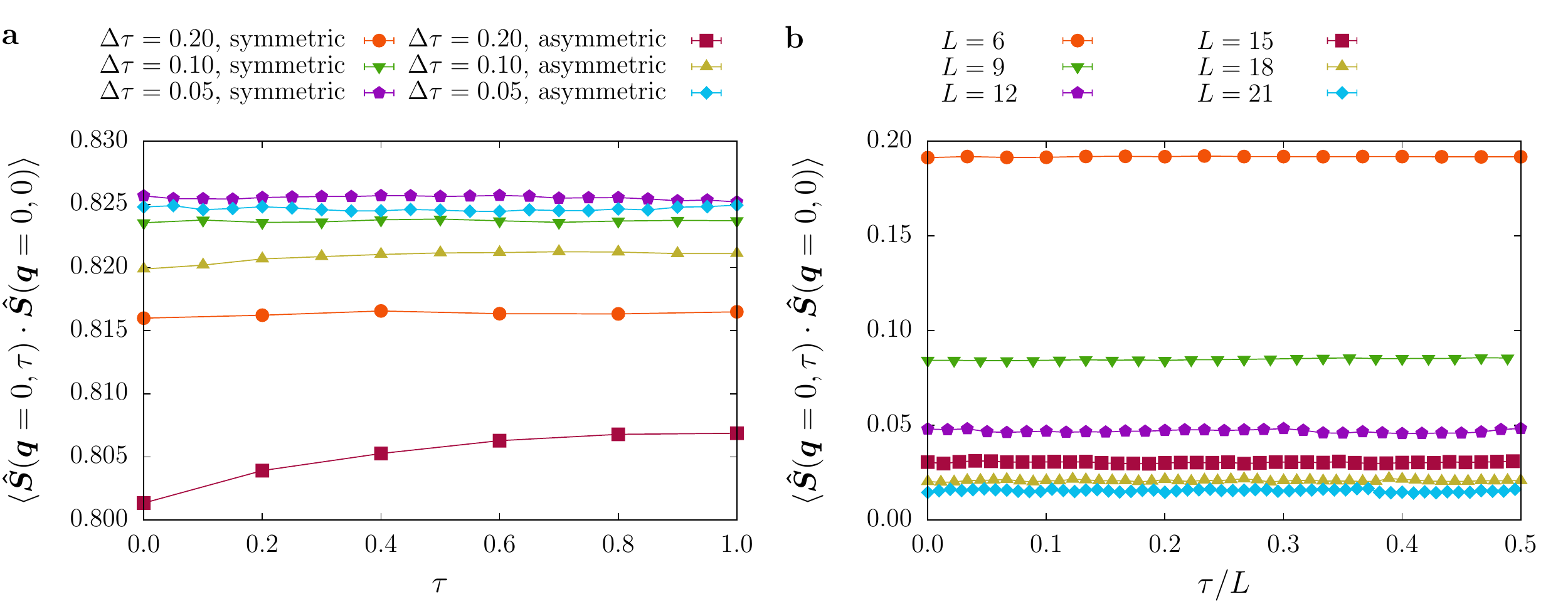}
\caption{\label{fig:Trotter}
{\bf a} Time-displaced spin correlation function at $\lambda=0.04$,
 $\beta=2$, and $L=3$. The label `symmetric' refers to the Trotter decomposition
 of equation~(\ref{Trotter}), whereas `asymmetric' refers to the alternative decomposition $ Z =
     \Tr \left[e^{- {\Delta \tau} \hat{H}_t}
   \left(\prod_{i=1}^{N}  e^{- {\Delta \tau} \hat{H}^x_\lambda (i) }  e^{- {\Delta \tau} \hat{H}^y_\lambda (i)}    e^{- {\Delta \tau} \hat{H}^z_\lambda (i) }
   \right)  \right]$.
{\bf b} Time-displaced spin correlation function for the symmetric
Trotter decomposition and $\lambda=0.019$, $\beta =L$.
   }
\end{figure}

\subsection{Trotter decomposition used for the QMC simulations}
\label{sec:Trotter}

For the finite-temperature QMC simulations underlying this work, imaginary
time was discretized with a spacing $\Delta \tau= \beta/L_\tau$. To ensure
hermiticity, the partition function is written as
\begin{equation}\label{Trotter}
 Z =
    \Tr \left[e^{- \frac{\Delta \tau}{2} \hat{H}_t} 
   \left(\prod_{i=1}^{N}  e^{-\frac{\Delta \tau}{2} \hat{H}^x_\lambda (i) }  e^{-\frac{\Delta \tau}{2} \hat{H}^y_\lambda (i)}    e^{- \frac{\Delta \tau}{2} \hat{H}^z_\lambda (i) }  \\
       \prod_{j=N}^{1}  e^{-\frac{\Delta \tau}{2} \hat{H}^z_\lambda (j) } e^{-\frac{\Delta \tau}{2} \hat{H}^y_\lambda (j)}  e^{- \frac{\Delta \tau}{2} \hat{H}^x_\lambda (j) } \right)
e^{ - \frac{\Delta \tau}{2} \hat{H}_t }  \right] ^{L_\tau}
\end{equation}
where $\hat{H}_t$ is defined by equation~(1) and the interaction~(2) was
partitioned into local operators $H^\alpha_\lambda(i)$ acting on spin
component $\alpha=x,y,z$ and on hexagon $i$. The leading discretization error
for the partition function then scales as $\Delta\tau^2$.

The Trotter decomposition in equation~(\ref{Trotter}) breaks the global SU(2) spin
rotation symmetry. For example, $[ \hat{H}^x_\lambda (i), \hat{H}^y_\lambda (i) ] \ne 0$,
so that equation~(\ref{Trotter}) will not be invariant under a global SU(2)
rotation. Because SU(2) symmetry breaking is a relevant perturbation for both
critical points considered, care has to be taken to ensure that its effects,
which scale as $\Delta \tau^2$, remain below the relevant energy scale. An
explicit test involves the total spin operator and generator of global SU(2) rotations
\begin{equation}
  \ve{S}_\text{tot}   =  \frac{1}{L} \sum_{\ve{r}, \ve{\delta} }  \ve{S}_{\ve{r},\ve{\delta}}\,.
\end{equation}
 Here,
$\ve{S}_{\ve{r},\ve{\delta}} =  \hat{\ve{c}}^{\dagger}_{\ve{r} + \ve{\delta}}
\ve{\sigma} \ve{c}^{}_{\ve{r} + \ve{\delta}}$ and $\ve{\delta}$ runs over the
positions of atoms in the unit cell at $\ve{r}$. S.Fig~\ref{fig:Trotter} shows
the associated time-displaced spin-spin correlation function. A global SU(2)
spin symmetry implies that this quantity is independent of imaginary time.
The numerical results are essentially constant in imaginary time if the
symmetric Trotter decomposition is used. Therefore, the latter was employed
together with $\Delta \tau = 0.2$ for all results of this work.

\begin{figure}
\centering
\includegraphics[width=0.675\textwidth]{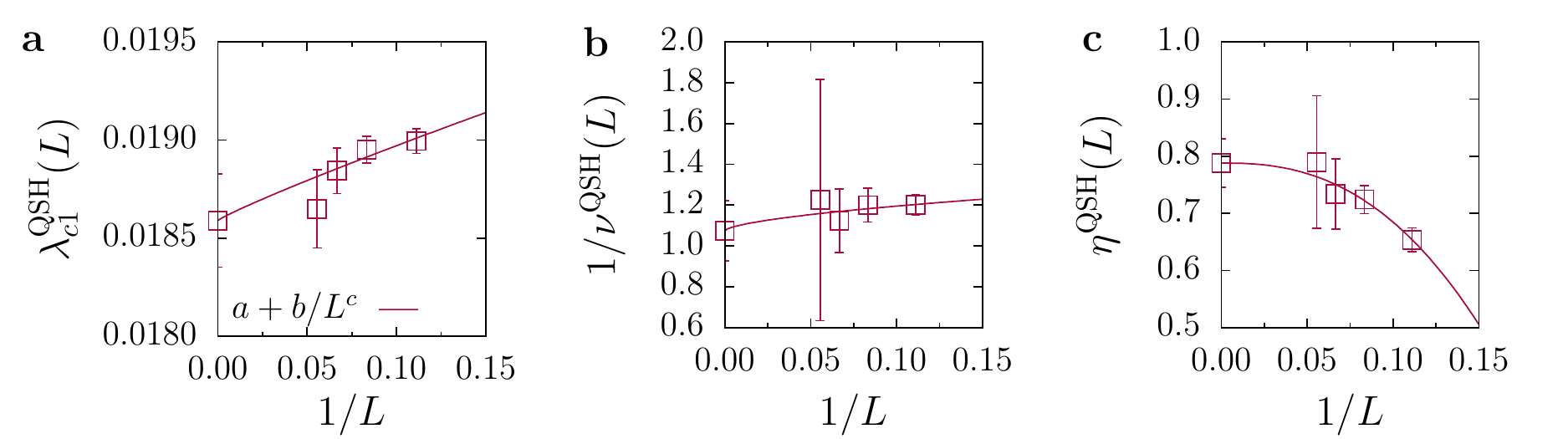}
  \caption{\label{fig:qshsm3}
    {\bf Gross-Neveu semimetal-QSH transition.}
    {\bf a} Extrapolation of the crossing points of $R^\text{QSH}_\chi$ for $L$
    and $L+3$ gives the critical value $\lambda^\text{QSH}_{c1}=0.0186(3)$. {\bf b} The inverse
    correlation length exponent $1/\nu^\text{QSH}=1.1(2)$. {\bf c} Estimation of the
    anomalous dimension  $\eta^\text{QSH}=0.79(5)$.}
\end{figure}

\subsection{Finite-size scaling analysis by different crossing points}

In addition to the main text, where a `two-size crossing' with sizes $L$ and
$L+6$ was performed, we provide a cross check based on crossings of $L$ and $L+3$.

\begin{figure}
\centering
\includegraphics[width=0.675\textwidth]{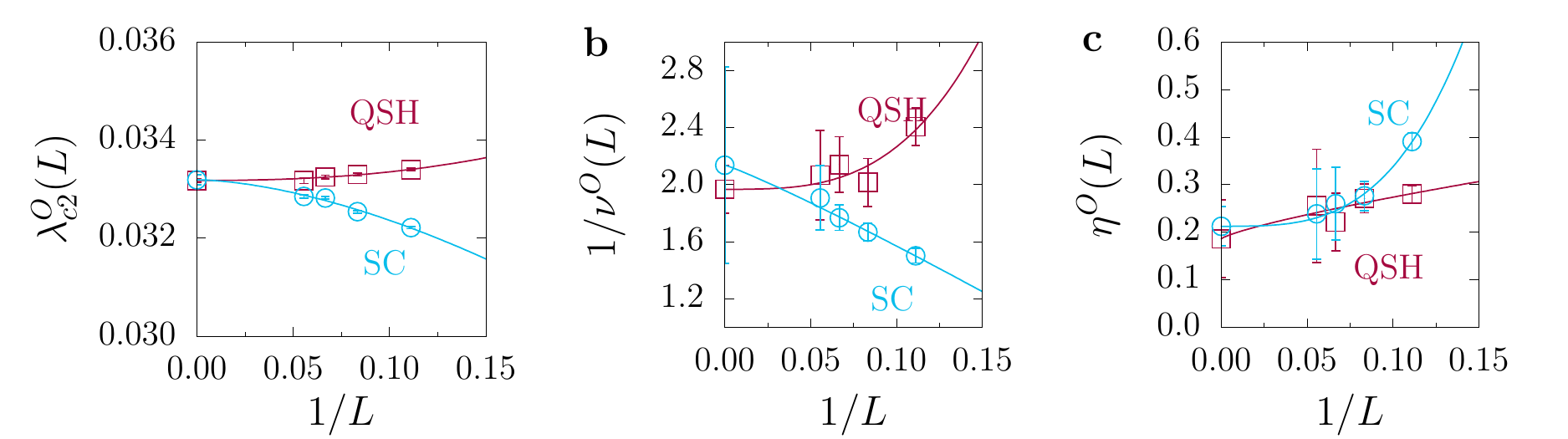}
  \caption{\label{fig:qshssc3}
    {\bf Deconfined QSH-SC transition.}
    {\bf a} Estimation of the critical values
    $\lambda_{c2}^\text{QSH}=0.03317(4)$ and
    $\lambda_{c2}^\text{SC}=0.0332(1)$. {\bf b},{\bf c} Critical
    exponents $1/\nu^\text{SC}=2.1(7)$, $1/\nu^\text{QSH}=2.0(2)$,
    $\eta^\text{SC}=0.21(5)$, and $\eta^\text{QSH}=0.19(9)$ from finite-size
    scaling of  the crossing points of $L$ and $L+3$.}
\end{figure}

When considering the $L$ and $L+6$ crossings in the main text for the estimation of $\lambda_{c2}^\text{SC}$ (see Fig.~3 of the main text), we are obliged
to take into account the $L=6$ data. Upon inspection, the fit turns out to be rather bad since  $\chi^2/\text{DOF}=66.9$.   On the other hand,
if we consider the $L$ and $L+3$ crossing points, we can omit the $L=6$ data
and get a more acceptable $\chi^2/\text{DOF}=6.8$. As apparent from
S.Fig.~\ref{fig:qshssc3}, the extrapolated value of $\lambda_{c2}^\text{SC}$
based on the crossing points of $L$ and $L+3$ compares favourably with the
analysis in the main text.

S.Fig~\ref{fig:qshsm3}\textbf{a},\textbf{b} show the crossing values of
{$\lambda$} and $1/ \nu$ at the semimetal-QSH transition, as obtained from the
correlation ratio. S.Fig~\ref{fig:qshsm3} \textbf{c} shows the anomalous
dimension $\eta$ at each crossing point, as well as a fit based on an expression
analogous to Eq.~(6) of the main text,
\begin{equation}\label{eq:etabycrossingpoints}
  \eta^{O}(L) = 2- \frac{1}{\log {r}} \left.\log \left( \frac{ \chi^{O}\left( \lambda, r L \right)  }{ \chi^{O}\left( \lambda,  L \right)  } \right)\right|_{\lambda = \lambda^O_c(L)}\,,
\end{equation}
where $r= \frac{L+3}{L}$.

The results of a similar analysis for the QSH-SC transition are reported in S.Fig.~\ref{fig:qshssc3}.

\subsection{Consistency of the finite-size scaling analysis}

As an independent consistency check on the results from the `two-size crossing'
method used in the main text, we consider in this section  a collective
fitting of multiple system sizes.

The collective fitting of $\lambda_{c}$ and $\nu$ is based on a polynomial
expansion of the scaling function of $R_{\chi}^O (L,\lambda)$  in
equation~(\ref{FSS}). Taking $\beta = L$, we have
\begin{equation}\label{eq:FSS_nv_expansion}
  R_{\chi}^O (L,\lambda)  \approx  \sum_{p=0}^{n}   a_p (\lambda - \lambda_c)^p L^{ p / \nu }  + L^{-\omega} \sum_{q=0}^{m} b_q  (\lambda - \lambda_c)^q L^{ q / \nu }\,.
\end{equation}
Here, $n$ and $m$ are the expansion orders for the dimensionless and the
scaling correction part of the universal function, respectively.

Table~\ref{table:DQCP_nv} reports the results of fits for the two order
parameters at the DQCP, including the case with $m=1$,  as well as the
case without considering any scaling correction.  We set $n=2$ for all the fits.  For
$m=1$,  the fitting of the QSH correlation ratio is satisfactory in terms of  $\chi^2
/\text{DOF}$  for $L_\text{min} \geq 9$, and the results are consistent
with each other for $L_\text{min}=9,12,15$.  Taking $\lambda_{c2} = 0.03314(5)$ and
$1/\nu=1.55(9)$ from the fit with $L_\text{min}=9$, we get consistency with
$\lambda^\text{QSH}_{c2}=0.03322(3)$ and $1 / \nu^\text{QSH}=1.7(4)$
from the `two-size crossing' analysis in the main text.
The results of a fit with a smaller data window are shown in the last three
rows of table~\ref{table:DQCP_nv}, revealing that the results are stable
upon variation of the number of degrees of freedom.
A fit using $R^\text{SC}_\chi$ also produces
acceptable values of $\chi^2 /\text{DOF}$  for $L_\text{min} =9$, $12$ and
$15$ and compares favourably with the results presented in the main text.

\begin{table}
 \begin{center}
  \begin{tabular}{| l | l | l | l| l | l|}
    \hline
    \multicolumn{6}{|c|}{DQCP -- QSH}                                                        \\ \hline
    $L_\text{min}$ & $\lambda_c$  &  $R^\text{QSH}_\chi(\lambda_c)$    &  $1/\nu$    & $\omega$ & $\chi^2 /\text{DOF}$ \\ \hline\hline
          9   &  0.03332(1) &  0.8665(4)  &   2.22(5)  &    n/a    &   139/44     \\ \hline
         12   &  0.03326(2) &  0.8702(8)  &   2.21(7)  &    n/a    &   58.9/32      \\ \hline
         15   &  0.03326(3) &  0.870(2)   &   2.3(2)   &    n/a    &   33.4/22      \\ \hline
         18   &  0.03321(5) &  0.875(5)   &   2.4(3)   &    n/a    &   13.9/12      \\ \hline \hline
          9   &  0.3314(5)  &  0.89(2)    &   1.55(9)  &  0.9(3)   &   29.1/41      \\ \hline
         12   &  0.331(2)   &  0.92(9)    &   1.5(2)   &  0.4(9)   &   24.7/29      \\ \hline
         15   &  0.331(2)   &  0.89(2)    &   1.6(3)   &  4(3)     &   15.1/19      \\ \hline \hline
          9   &  0.3315(5)  &  0.89(2)    &   1.5(2)   &  0.9(4)   &   28.2/36      \\ \hline
         12   &  0.331(2)   &  0.91(8)    &   1.5(2)   &  0.5(9)   &   23.8/26      \\ \hline
         15   &  0.331(2)   &  0.89(3)    &   1.7(3)   &  3(3)     &   13.9/16      \\ \hline
  \end{tabular}
\quad
  \begin{tabular}{| l | l | l | l| l | l|}
    \hline
    \multicolumn{6}{|c|}{DQCP -- SC}  \\ \hline
    $L_\text{min}$ & $\lambda_c$  &  $R^\text{SC}_\chi(\lambda_c)$    &  $1/\nu$   & $\omega$ & $\chi^2 /\text{DOF}$ \\ \hline\hline
          9   &  0.032675(4) &  0.8592(4) &   1.32(3)  &     n/a    &   1239/40       \\ \hline
         12   &  0.032791(5) &  0.8742(5) &   1.60(4)  &     n/a    &   123/29       \\ \hline
         15   &  0.032843(8) &  0.882(1)  &   1.83(6)  &     n/a    &   15.9/18      \\ \hline
         18   &  0.03286(3)  &  0.884(4)  &   1.9(2)   &     n/a    &   2.23/9       \\ \hline \hline
          9   &  0.03296(5)  &  0.907(9)  &   2.1(2)   &  1.7(3)    &   53.8/37     \\ \hline
         12   &  0.03287(2)  &  0.887(2)  &   2.3(2)   &  4.4(8)    &   26.0/26     \\ \hline
         15   &  0.0329(2)   &  0.89(3)   &   2.0(4)   &  3(13)     &   15.1/15      \\ \hline \hline
          9   &  0.0331(2)   &  0.94(4)   &   1.9(2)   &  1.1(5)    &   42.5/33     \\ \hline
         12   &  0.03285(3)  &  0.884(5)  &   2.0(3)   &  8(6)      &   20.6/22     \\ \hline
         15   &  0.0329(2)   &  0.89(3)   &   1.7(5)   &  4(16)     &   14.4/13      \\ \hline
  \end{tabular}
\end{center}
\caption{ \label{table:DQCP_nv}
  Collective fitting at the DQCP for $\lambda_c$, $1 / \nu$, and $R^O_\chi(\lambda_c)$
  based on the correlation ratios $R_{\chi}^O(L,\lambda) $ and equation~(\ref{eq:FSS_nv_expansion}). We compare the
  case without taking into account scaling corrections (data rows 1--4) to the case with
  $m=1$ (rows 5--7 correspond to a larger data window, rows 8--10 to a smaller one); in both cases, $n=2$.}
\end{table}

To reduce the number of degrees of freedom in the fit of the anomalous dimension $\eta$, a substitution in terms of the  scaling form
for $\chi^O $ and $R_{\chi}^O  $ in equation~(\ref{FSS}) is performed, using
the expansion (we ignore the correction-to-scaling term)
 \begin{equation}\label{eq:FSS_eta_expansion}
 \begin{aligned}
  \chi^O (L, R) & = L^{2-\eta} f (R)     \approx  L^{2-\eta} \sum_{p=0}^n  a_p  R^p\,.
 \end{aligned}
 \end{equation}
 As shown in table~\ref{table:DQCP_eta}, the collective fitting for both
 order parameters is acceptable for $L_\text{min} \geq 12$.   The corresponding values $\eta^\text{QSH} = 0.194(9)$ and $\eta^\text{SC} = 0.279(9)$
 agree well with the values $0.21(5)$ and $0.22(6)$ obtained with the `two-size crossing' approach  used in the main text.

 \begin{table}
 \begin{center}
  \begin{tabular}{| l | l | l |l|}
    \hline
    \multicolumn{3}{|c|}{DQCP -- QSH}                                                        \\ \hline
    $L_\text{min}$ &  $\eta$     &  $\chi^2 /\text{DOF}$                     \\ \hline\hline
          6   &  0.30(2)    &  990/68                          \\ \hline
          9   &  0.24(1)    &  249/57                          \\ \hline
          12  &  0.19(2)    &  92.0/40                           \\ \hline
          15  &  0.24(3)    &  49.8/25                           \\ \hline
          18  &  0.18(6)    &  12.0/12                           \\ \hline
  \end{tabular}\hspace*{2em}
  \begin{tabular}{| l | l | l |l|}
    \hline
    \multicolumn{3}{|c|}{DQCP -- SC\vphantom{Q}}                        \\ \hline
    $L_\text{min}$ &  $\eta$     &  $\chi^2 /\text{DOF}$                     \\ \hline\hline
          6   &  0.344(8)   &  525/45                          \\ \hline
          9   &  0.308(7)   &  128/36                          \\ \hline
          12  &  0.28(2)    &  65.6/26                           \\ \hline
          15  &  0.23(2)    &  24.2/17                           \\ \hline
          18  &  0.19(9)    &  7.47/8                           \\ \hline
  \end{tabular}
\end{center}
\caption{\label{table:DQCP_eta}
Same analysis as in table~\ref{table:DQCP_nv} but for the exponent $\eta$
using equation~(\ref{eq:FSS_eta_expansion}). Here $n=2$, while scaling corrections are ignored.
}
\end{table}

 \begin{table}
 \begin{center}
  \begin{tabular}{| l | l | l | l| l | l|}
    \hline
    \multicolumn{6}{|c|}{Gross-Neveu -- QSH}                                                                 \\ \hline
    $L_\text{min}$ & $\lambda_{c}$  &  $R^\text{QSH}_\chi(\lambda_c)$     &  $1/\nu$    & $\omega$ & $\chi^2 /\text{DOF}$  \\ \hline\hline
          6   &  0.01898(2) &  0.6970(6)  &   1.26(2)   &     n/a\hspace*{2em}    &   94.6/34      \\ \hline
          9   &  0.01891(2) &  0.693(1)   &   1.17(3)   &     n/a    &   34.8/27      \\ \hline
         12   &  0.01882(4) &  0.687(2)   &   1.17(5)   &     n/a    &   18.6/20      \\ \hline
         15   &  0.01870(7) &  0.678(5)   &   1.14(11)  &     n/a    &   9.22/13      \\ \hline
         18   &  0.0186(2)  &  0.67(2)    &   1.3(4)    &     n/a    &   1.90/6       \\ \hline
  \end{tabular}
  \hspace*{2em}
    \begin{tabular}{| l | l | l | l|}
    \hline
    \multicolumn{3}{|c|}{Gross-Neveu -- QSH} \\ \hline
    $L_\text{min}$ &  $\eta$     &  $\chi^2 /\text{DOF}$                     \\ \hline\hline
          6   &  0.666(7)   &  210/33                          \\ \hline
          9   &  0.70(1)    &  111/26                           \\ \hline
          12  &  0.76(2)    &  28.6/18                            \\ \hline
          15  &  0.78(2)    &  4.4/12                            \\ \hline
          18  &  0.81(6)    &  1.9/6                              \\ \hline
  \end{tabular}
\end{center}
\caption{\label{table:GNY}
Same as table~\ref{table:DQCP_nv}, but for the Gross-Neveu transition.
Here $n=2$, while scaling corrections are ignored.
}
\end{table}

We also carried out the  collective fitting  at the Gross-Neveu critical point.   In contrast to the  DQCP, this phase transition suffers
much less from  corrections  to scaling (as shown in Fig.~3 of the main text, the crossing points converge quickly).
Hence,  a fit without the scaling correction term is performed,  and  the results  are shown in table~\ref{table:GNY}.
As can be seen, a good $\chi^2 /\text{DOF}$ is obtained once the $L=6$ data set
is neglected, and fits for $L_{min} = 9,12,15$ or $18$ produce consistent
results. Taking $\lambda_{c1} = 0.01891(2)$  and $1/\nu = 1.17(3)$ from
$L_\text{min}=9$, the results match those of the analysis in the main text.

Results for the anomalous dimension at the Gross-Neveu critical point are
listed in table~\ref{table:GNY}. The fits yield acceptable $\chi^2$ values for
$L_\text{min} \geq 12 $. The exponent $\eta = 0.76(1) $ from $L_\text{min} = 12$ also matches the analysis in the main text.


 \subsection{Single-particle gap and free-energy derivative across the QSH-SC transition}

\begin{figure}[b]
\centering
\includegraphics[width=0.9\textwidth]{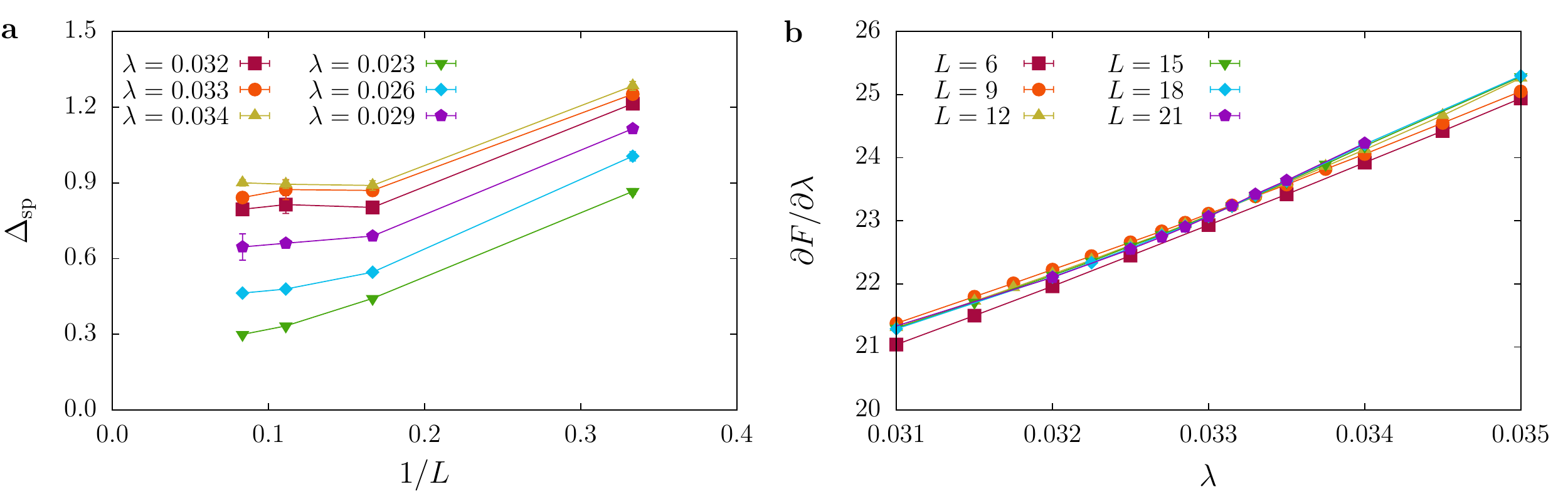}
\caption{\label{fig:spgap}
 {\bf a} Fermionic single-particle gap and {\bf b} free-energy derivative
 $\partial F /  \partial {\lambda}$ across the QSH-SC transition  at    $\lambda_{c2}^\text{SC}=0.0331(3)$.
}
\end{figure}

The single-particle gap $\Delta_\text{sp}$ is obtained from the
single-particle Green function
\begin{equation}
  G (\ve{k}, \tau)  =  \frac{1}{L^2} \sum_{\ve{r},\ve{r'},\ve{\delta},\sigma}  \langle  \hat{c}^{\dagger}_{\ve{r}+\ve{\delta},\sigma} (\tau) \hat{c}^{}_{\ve{r} +\ve{\delta},\sigma}( 0) \rangle   e^{ \mathrm{i} \ve{k} \cdot (\ve{r} - \ve{r'}) }
\end{equation}
where $\ve{r}+\ve{\delta} $ runs over the two orbitals of the unit cell located at $\ve{r}$.
The single-particle gap is minimal at the Dirac point $\mathrm{K} = (\frac{4\pi}{3}, 0 )$ and is extracted  by noting that asymptotically
\begin{equation}
 G(\mathrm{K}, \tau)  \propto e^{-\Delta_\text{sp} \tau }\,.
\end{equation}
Here, we used $\beta = 36$. S.Fig~\ref{fig:spgap}a demonstrates that $\Delta_\text{sp}$
remains nonzero across the QSH-SC transition at $\lambda_{c2} \approx 0.033$.

In order to clarify nature of the QSH-SC transition, we also calculated the
first partial derivative of the free energy  density with respect to the
coupling $\lambda$ (we use the same notation as in the main text)
\begin{equation}
\frac{\partial F}{\partial\lambda} =
- \frac{1}{L^2} \sum_{\hexagon}   \left\< \left( \sum_{ \langle \langle \ve{i},\ve{j} \rangle \rangle  \in \hexagon } \mathrm{i}  \nu_{\bm{ij}}
 \hat{\boldsymbol{c}}^{\dagger}_{\ve{i}} \boldsymbol{\sigma} \hat{\boldsymbol{c}}^{}_{\ve{j}}+\text{H.c.} \right)^2 \right\>.
\label{F-over-lambda}
\end{equation}
S.Fig~\ref{fig:spgap}b shows $\partial F / \partial \lambda$ for $\beta=L$
in the vicinity of $\lambda_{c2} \approx 0.033$.   As  expected for a
continuous transition, we observe no sign of a jump.

\subsection{Charged skyrmion defects of the QSH state}

In Ref.~\cite{Grover08}, it was shown that when a QSH state is generated by
spontaneous symmetry breaking,  skyrmion defects of the vector order
parameter will carry an electric charge of $Q_\text{e}=2\mathrm{e}$, leading
to the relation
\begin{equation}
\label{C-to-Q}
  Q_\text{e} = 2 \text{e} Q
\end{equation}
where $Q$ is the Pontryagin index that counts  the  winding of the unit vector order parameter on the sphere.

Here, we substantiate this fact in terms of an explicit calculation for a
lattice model. Our starting point is the Hamiltonian
\begin{equation}
\label{H-Ski}
\hat{H} = -t \sum_{ \langle  \ve{i}  \ve{j} \rangle }  \left( \hat{\ve{c}}^{\dagger}_{ \ve{i}}  \ve{\hat{c}}^{}_{ \ve{j} } + \text{H.c.}  \right)
 +  \lambda \sum_{\hexagon}  \ve{N}(\ve{x}) \cdot    \Bigg(\sum_{\langle \langle \ve{i},\ve{j} \rangle \rangle \in {\hexagon} }
  \underbrace{ \mathrm{i} \nu_{\ve{i}\ve{j}}
  \ve{\hat{c}}^{\dagger}_{\ve{i}}  \ve{\sigma} \ve{\hat{c}}^{}_{\ve{j}}    + \text{H.c.}  }_{\equiv \ve{\hat{J}}_{\ve{i}\ve{j}}}\Bigg)\,,
\end{equation}
where $\ve{N}(\ve{x})=(N^x(\ve{x}), N^y(\ve{x}), N^z(\ve{x}))$ is a  unit vector at position  $\ve{x}$ corresponding to  the centre of a hexagon.
Since $\hat{H}$ is invariant under  time reversal symmetry,
$\hat{T}^{-1} \alpha \binom{ \hat{c}_{i,\uparrow}}{ \hat{c}_{i,\downarrow} }  \hat{T} =
  \overline{\alpha} \binom { \hat{c}_{i,\downarrow} } {-  \hat{c}_{i,\uparrow}}  $, Kramers' theorem holds and  stipulates that all
  eigenstates are doubly degenerate.

On the honeycomb lattice, the Pontryagin index is defined as
\begin{equation}\label{Q-lattice}
 Q = \frac{1}{8 \pi} \sum_{\ve{ x}}  \ve {N}(\ve{x})  \cdot (\ve {N} (\ve {x }+ \ve{a}_1) - \ve{ N }(\ve{x}) ) \times
 [(\ve{N} (\ve{ x}) - \ve{ N }(\ve{x} + \ve{a}_2) ) + ( \ve{ N} (\ve{ x}) - \ve{N} (\ve{ x} - \ve{ a}_1 + \ve{ a}_2) )]
\end{equation}
with unit vectors  $ \ve{a}_1 = (1,0) $ and $ \ve{a}_2 = (\frac{1}{2},\frac{\sqrt{3}}{2}) $.

For an arbitrary vector field $\ve{N}  (\ve{x})$,  Hamiltonian~(\ref{H-Ski}) does not preserve particle-hole (P-H) symmetry.  For example,
defining the P-H transformation as
\begin{equation}
\hat{P}_{z}^{-1} \alpha \binom{\hat{c}^{\dagger}_{\ve{i},\uparrow}}{ \hat{c}^{\dagger}_{\ve{i},\downarrow} }  \hat{P}_z =
 {\eta _i} \overline{\alpha} \binom {\phantom{-} \hat{c}_{\ve{i},\uparrow} } {- \hat{c}_{\ve{i},\downarrow}}\,,
\end{equation}
where $\eta_i = 1\, (-1)$ for $i \in \mathrm{A}\,(\mathrm{B})$,  we have
\begin{equation}
 \begin{aligned}
 &  \hat{P}^{-1}_{z} \hat{J} ^x_{\ve{i},\ve{j}}\hat{P}_z  = \phantom{-}\hat{J}^x_{\ve{i},\ve{j}}\,,   \\
 &  \hat{P}^{-1}_{z} \hat{J}^y_{\ve{i},\ve{j}} \hat{P}_z  = \phantom{-} \hat{J}^y_{\ve{i},\ve{j}}\,,   \\
 &  \hat{P}^{-1}_{z} \hat{J}^z _{\ve{i},\ve{j}}\hat{P}_z  = -\hat{J}^z_{\ve{i},\ve{j}}\,.
 \end{aligned}
\end{equation}

\begin{figure}[t]
\centering     
\includegraphics[width=0.72\textwidth]{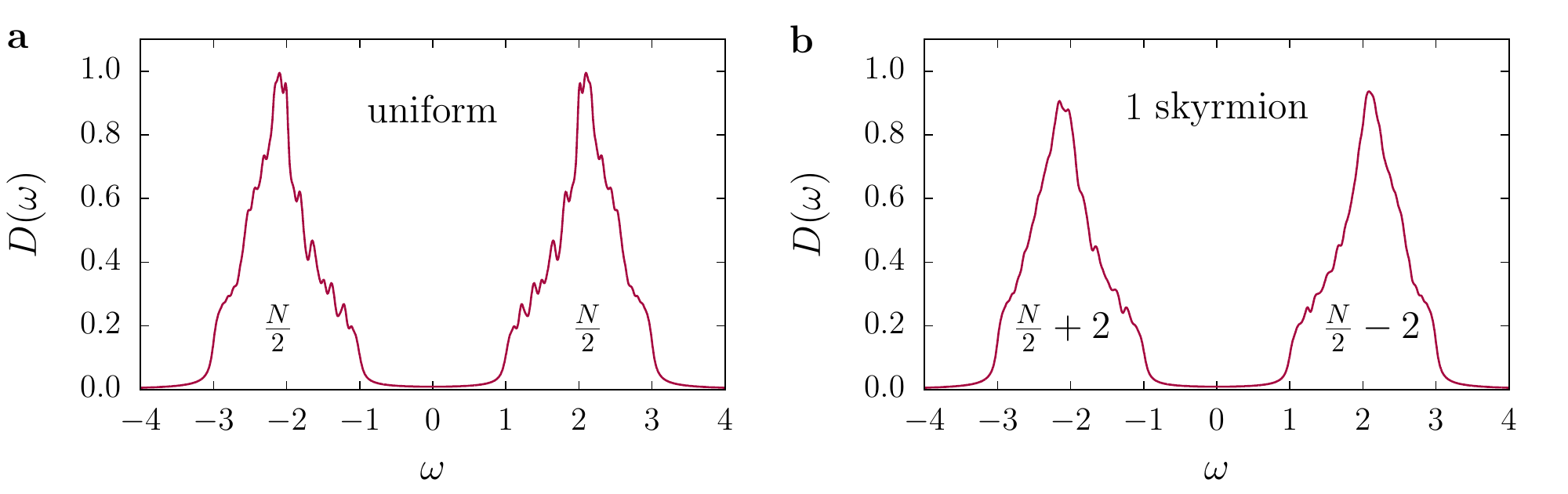}
\caption{\label{fig:Nw-M-QSH}
Density of states $N(\omega)$ of Hamiltonian~(\ref{H-Ski}) for {\bf a} uniform polarisation, {\bf b} a `single skyrmion' configuration with
$Q\approx-0.989$. We have included an artificial broadening by using the form
$D(\omega) = - \pi^{-1} \sum_n \mathrm{Im}\,(\omega - \varepsilon _n +
\mathrm{i} \delta )^{-1} $, where $\varepsilon _n$ are the eigenvalues and $\delta=0.05$. Here, $L=36$, $\lambda=0.5$.
}
\end{figure}

A general P-H transformation can be written as
\begin{equation}
  \hat{P}(\theta, \phi)  =   \hat{U}^{-1} (\theta, \phi )  \hat{P}_z  \hat{U} (\theta , \phi )
\end{equation}
where
\begin{equation}
\hat{U}^{-1} (\theta, \phi) \binom {\hat{c}_{i, \uparrow}}  { \hat{c}_{i, \downarrow} } \hat{U} (\theta, \phi)  = \begin{pmatrix} \cos(\theta/2)  & - \sin(\theta/2)  e^{-i \phi}
\\  \sin(\theta/2) e^{i \phi}  &  \cos(\theta/2) \end{pmatrix} \binom {\hat{c}_{\ve{i},
  \uparrow}}  { \hat{c}_{\ve{i}, \downarrow} }.
\end{equation}
For Hamiltonian~(\ref{H-Ski}), it yields
\begin{equation}
  \hat{P}^{-1}(\theta, \phi) \hat{H} (\ve N) \hat{P}(\theta, \phi) = H ( \ve {N}' )
\end{equation}
where
\begin{equation}
  \ve{N}' (\ve x) = R^{-1} (\theta, \phi)  \begin{pmatrix}  1  & 0 & 0 \\ 0 & 1 & 0 \\ 0 & 0 & -1 \\ \end{pmatrix} R (\theta, \phi) \ve N ({\ve x})
\end{equation}
and
\begin{equation}
  R(\theta,\phi) = \begin{pmatrix}  \cos^2 ({\theta}/{2}) - \sin^2 ({\theta}/{2}) \cos (2 \phi)  &
                                   -\sin^2 ({\theta}/{2}) \sin(2\phi)                               &
                                   -\sin(\theta) \cos(\phi)                                \\
                                   -\sin^2 ({\theta}/{2}) \sin(2\phi)                               &
                                    \cos^2 ({\theta}/{2}) + \sin^2 ({\theta}/{2}) \cos (2 \phi)  &
                                   - \sin(\theta) \sin(\phi)                                \\
                                     \sin(\theta) \cos(\phi)                                        &
                                     \sin(\theta) \sin(\phi)                                      &
                                     \cos(\theta)                                           \\
  \end{pmatrix}.
\end{equation}

Thus, there is no generic P-H transformation that leaves this Hamiltonian invariant, unless $\ve {N}(\ve{ x} )$ is varied in an $R^2$ space
($\theta$ and $\phi$ can be defined such that $\sin \theta \cos \phi N_x + \sin \theta \sin \phi N_y + \cos \theta N_z =0$). Since the transformation
has a determinant of $-1$, the generic P-H transformation gives
\begin{equation}
  Q(\ve {N}' (\ve{ x})) = - Q(\ve{ N }(\ve{ x}) ).
\end{equation}
The `electric charge' refers to the  number of occupied states at zero temperature, relative to half filling. The
sign change of the  Pontryagin index under a  P-H transformation provides a natural way of understanding equation~(\ref{C-to-Q}).
Note that in contrast to a skyrmion,  a 2D  topological defect (such as vortex) has a vanishing  Pontryagin index  and carries no charge.

The argument of charged skyrmions  fails  when other  Dirac mass terms are considered. For example, a system with fluctuations of a
three component vector field Yukawa-coupled to the three  antiferromagnetic mass terms  does not break P-H symmetry.   In this case, a
skyrmion configuration with nonzero Pontryagin index does not carry electric charge.

We diagonalised Hamiltonian~(\ref{H-Ski}) on a honeycomb lattice with $L=36$,
setting $t=1$ and $\lambda=0.5$. S.Fig~\ref{fig:Nw-M-QSH} compares the density of states for a uniform field
$\ve{N} (\ve{ x} )$ and for a `hedgehog' configuration corresponding to a single  skyrmion.
On the lattice,  the Pontryagin index is not quantised and we obtain $ Q \approx -0.989$.
The system remains gapped when one skyrmion is inserted, see S.Fig.~\ref{fig:Nw-M-QSH}b.   The breaking of P-H
symmetry is also apparent from  $D(\omega) \ne D(-\omega) $. Simple number counting shows that
\begin{equation}
\int^{0}_{-\infty} D(\omega) d\omega = N/2 + 2 \,,\quad \int^{\infty}_{0} D(\omega) d\omega = N/2 - 2.
\end{equation}
Compared to the case of uniform polarisation in S.Fig.~\ref{fig:Nw-M-QSH}a, an additional charge 2e is generated.

\begin{figure}
\centering
\includegraphics[width=0.36\textwidth]{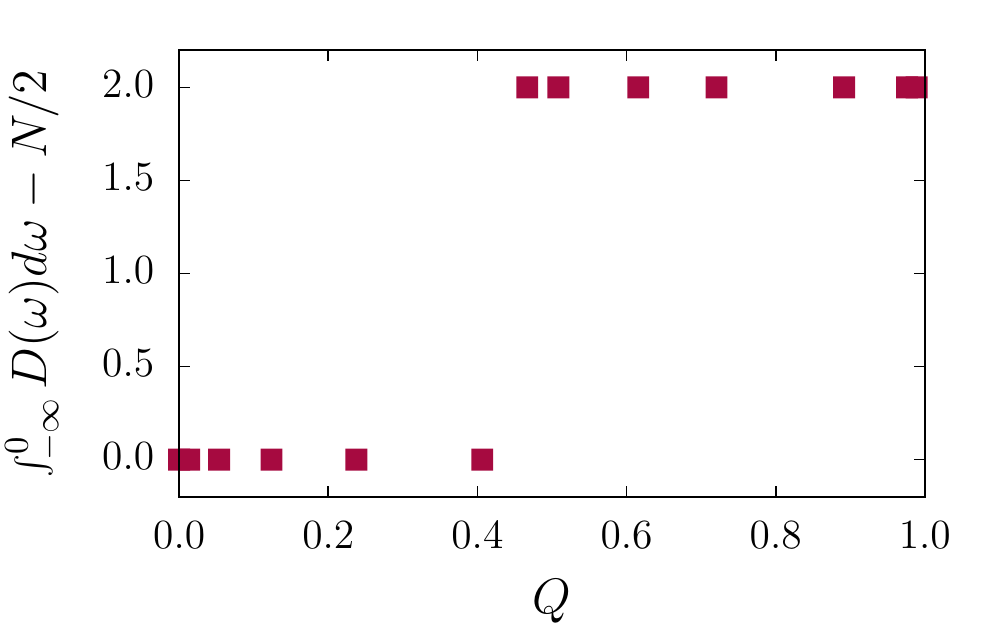}
\caption{\label{fig:Pump}
Integrated density of states as a function of $Q$ for open boundary
conditions and $L=36$.}
\end{figure}

\begin{figure}
\centering
\includegraphics[width=0.72\textwidth]{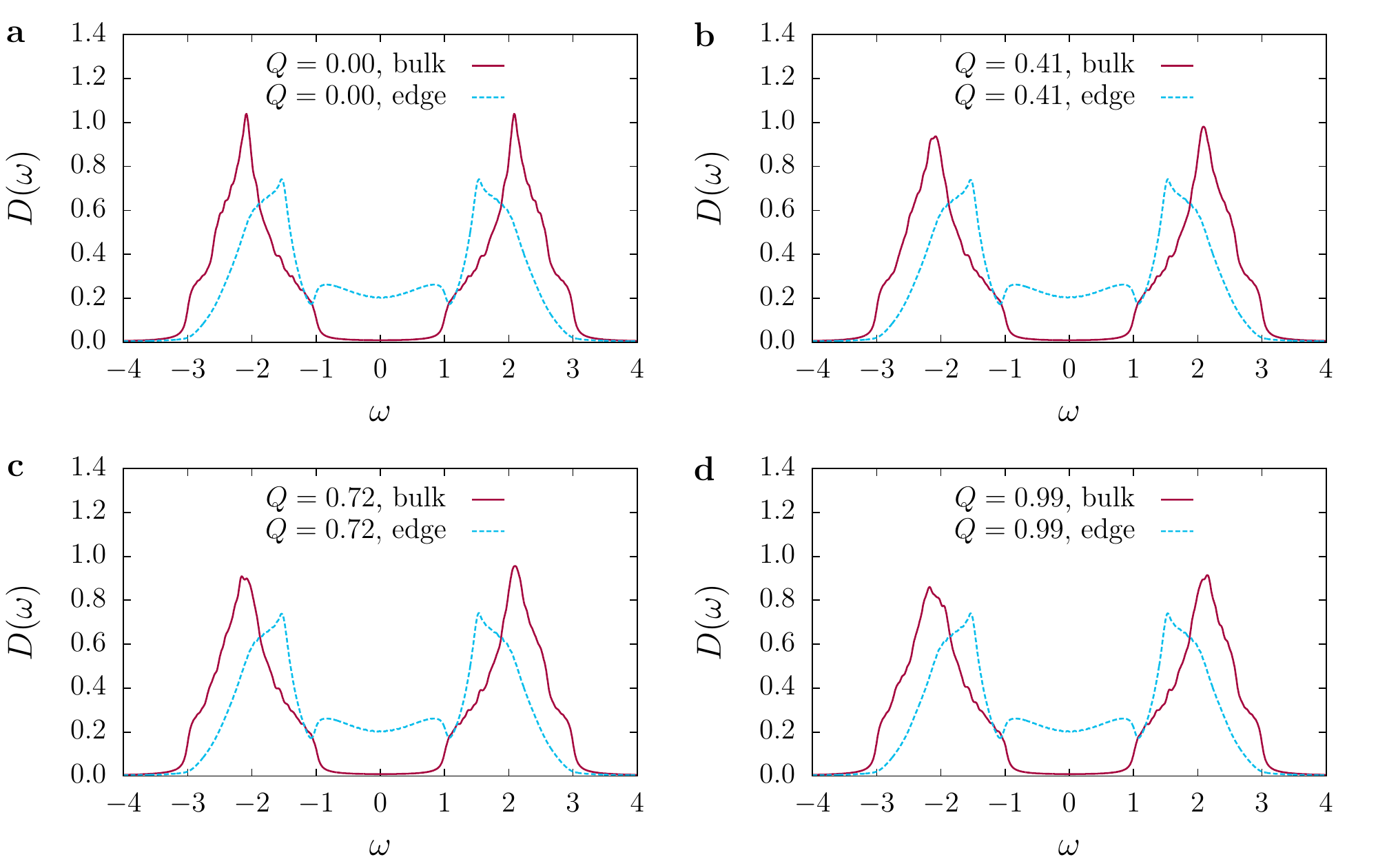}
\caption{\label{fig:Edge-Bulk}
Edge and bulk density of states for different $\ve{N} (\ve{x})$,
corresponding to the values of $Q$ given in each panel. Here, $L=36$, and $D(\omega)$ was
broadened as explained in S.Fig.~\ref{fig:Nw-M-QSH}.
}
\end{figure}

On a system with open boundary conditions,  the  Pontryagin index  is not necessarily quantised and we can investigate  how charge is
 transferred during the  {\textit{insertion}} of a skyrmion by varying the Pontryagin index from zero to one.
S.Fig~\ref{fig:Pump} shows that the total charge is `pumped' from $0$ to 2e
and we observe a step function at a non-integer  value of   $Q$.
This is a  consequence of the aforementioned Kramers theorem.  As shown in S.Fig.~\ref{fig:Edge-Bulk},
the bulk remains gapped during this process, while the edge stays gapless.  Thus,  the charge 2e  is pumped through the edge under
insertion of a skyrmion.

\begin{thebibliography}{10}
\expandafter\ifx\csname url\endcsname\relax
  \def\url#1{\texttt{#1}}\fi
\expandafter\ifx\csname urlprefix\endcsname\relax\def\urlprefix{URL }\fi
\providecommand{\bibinfo}[2]{#2}
\providecommand{\eprint}[2][]{\url{#2}}

\bibitem{KaneMele05b}
\bibinfo{author}{Kane, C.~L.} \& \bibinfo{author}{Mele, E.~J.}
\newblock ${Z}_{2}$ Topological Order and the Quantum Spin Hall Effect.
\newblock \emph{\bibinfo{journal}{Phys. Rev. Lett.}}
  \textbf{\bibinfo{volume}{95}}, \bibinfo{pages}{146802}
  (\bibinfo{year}{2005}).

\bibitem{Raghu08}
\bibinfo{author}{Raghu, S.}, \bibinfo{author}{Qi, X.-L.},
  \bibinfo{author}{Honerkamp, C.} \& \bibinfo{author}{Zhang, S.-C.}
\newblock Topological Mott Insulators.
\newblock \emph{\bibinfo{journal}{Phys. Rev. Lett.}}
  \textbf{\bibinfo{volume}{100}}, \bibinfo{pages}{156401}
  (\bibinfo{year}{2008}).

\bibitem{Koenig07}
\bibinfo{author}{K\"onig, M.} \emph{et~al.}
\newblock Quantum Spin Hall Insulator State in HgTe Quantum Wells.
\newblock \emph{\bibinfo{journal}{Science}} \textbf{\bibinfo{volume}{318}},
  \bibinfo{pages}{766} (\bibinfo{year}{2007}).

\bibitem{Reis17}
\bibinfo{author}{Reis, F.} \emph{et~al.}
\newblock Bismuthene on a SiC substrate: A candidate for a high-temperature
  quantum spin Hall material.
\newblock \emph{\bibinfo{journal}{Science}}  (\bibinfo{year}{2017}).

\bibitem{Gross74}
\bibinfo{author}{Gross, D.~J.} \& \bibinfo{author}{Neveu, A.}
\newblock Dynamical symmetry breaking in asymptotically free field theories.
\newblock \emph{\bibinfo{journal}{Phys. Rev. D}} \textbf{\bibinfo{volume}{10}},
  \bibinfo{pages}{3235--3253} (\bibinfo{year}{1974}).

\bibitem{Grover08}
\bibinfo{author}{Grover, T.} \& \bibinfo{author}{Senthil, T.}
\newblock Topological Spin Hall States, Charged Skyrmions, and
  Superconductivity in Two Dimensions.
\newblock \emph{\bibinfo{journal}{Phys. Rev. Lett.}}
  \textbf{\bibinfo{volume}{100}}, \bibinfo{pages}{156804}
  (\bibinfo{year}{2008}).

\bibitem{Senthil04_1}
\bibinfo{author}{Senthil, T.}, \bibinfo{author}{Balents, L.},
  \bibinfo{author}{Sachdev, S.}, \bibinfo{author}{Vishwanath, A.} \&
  \bibinfo{author}{Fisher, M. P.~A.}
\newblock Quantum criticality beyond the Landau-Ginzburg-Wilson paradigm.
\newblock \emph{\bibinfo{journal}{Phys. Rev. B}} \textbf{\bibinfo{volume}{70}},
  \bibinfo{pages}{144407} (\bibinfo{year}{2004}).

\bibitem{Senthil04_2}
\bibinfo{author}{Senthil, T.}, \bibinfo{author}{Vishwanath, A.},
  \bibinfo{author}{Balents, L.}, \bibinfo{author}{Sachdev, S.} \&
  \bibinfo{author}{Fisher, M. P.~A.}
\newblock Deconfined Quantum Critical Points.
\newblock \emph{\bibinfo{journal}{Science}} \textbf{\bibinfo{volume}{303}},
  \bibinfo{pages}{1490--1494} (\bibinfo{year}{2004}).

\bibitem{WangC17}
\bibinfo{author}{Wang, C.}, \bibinfo{author}{Nahum, A.},
  \bibinfo{author}{Metlitski, M.~A.}, \bibinfo{author}{Xu, C.} \&
  \bibinfo{author}{Senthil, T.}
\newblock Deconfined Quantum Critical Points: Symmetries and Dualities.
\newblock \emph{\bibinfo{journal}{Phys. Rev. X}} \textbf{\bibinfo{volume}{7}},
  \bibinfo{pages}{031051} (\bibinfo{year}{2017}).

\bibitem{Shao15}
\bibinfo{author}{Shao, H.}, \bibinfo{author}{Guo, W.} \&
  \bibinfo{author}{Sandvik, A.~W.}
\newblock Quantum criticality with two length scales.
\newblock \emph{\bibinfo{journal}{Science}} \textbf{\bibinfo{volume}{352}},
  \bibinfo{pages}{213--216} (\bibinfo{year}{2016}).

\bibitem{Nahum15}
\bibinfo{author}{Nahum, A.}, \bibinfo{author}{Chalker, J.~T.},
  \bibinfo{author}{Serna, P.}, \bibinfo{author}{Ortu\~no, M.} \&
  \bibinfo{author}{Somoza, A.~M.}
\newblock Deconfined Quantum Criticality, Scaling Violations, and Classical
  Loop Models.
\newblock \emph{\bibinfo{journal}{Phys. Rev. X}} \textbf{\bibinfo{volume}{5}},
  \bibinfo{pages}{041048} (\bibinfo{year}{2015}).

\bibitem{Nahum15_1}
\bibinfo{author}{Nahum, A.}, \bibinfo{author}{Serna, P.},
  \bibinfo{author}{Chalker, J.~T.}, \bibinfo{author}{Ortu\~no, M.} \&
  \bibinfo{author}{Somoza, A.~M.}
\newblock Emergent SO(5) Symmetry at the N\'eel to Valence-Bond-Solid
  Transition.
\newblock \emph{\bibinfo{journal}{Phys. Rev. Lett.}}
  \textbf{\bibinfo{volume}{115}}, \bibinfo{pages}{267203}
  (\bibinfo{year}{2015}).

\bibitem{Kawashima07}
\bibinfo{author}{Kawashima, N.} \& \bibinfo{author}{Tanabe, Y.}
\newblock Ground States of the $\mathrm{SU}(N)$ Heisenberg Model.
\newblock \emph{\bibinfo{journal}{Phys. Rev. Lett.}}
  \textbf{\bibinfo{volume}{98}}, \bibinfo{pages}{057202}
  (\bibinfo{year}{2007}).

\bibitem{Sandvik07}
\bibinfo{author}{Sandvik, A.~W.}
\newblock Evidence for Deconfined Quantum Criticality in a Two-Dimensional
  Heisenberg Model with Four-Spin Interactions.
\newblock \emph{\bibinfo{journal}{Phys. Rev. Lett.}}
  \textbf{\bibinfo{volume}{98}}, \bibinfo{pages}{227202}
  (\bibinfo{year}{2007}).

\bibitem{Berg12}
\bibinfo{author}{Berg, E.}, \bibinfo{author}{Metlitski, M.~A.} \&
  \bibinfo{author}{Sachdev, S.}
\newblock Sign-Problem--Free Quantum Monte Carlo of the Onset of
  Antiferromagnetism in Metals.
\newblock \emph{\bibinfo{journal}{Science}} \textbf{\bibinfo{volume}{338}},
  \bibinfo{pages}{1606--1609} (\bibinfo{year}{2012}).

\bibitem{Kaul13}
\bibinfo{author}{Kaul, R.~K.}, \bibinfo{author}{Melko, R.~G.} \&
  \bibinfo{author}{Sandvik, A.~W.}
\newblock Bridging Lattice-Scale Physics and Continuum Field Theory with
  Quantum Monte Carlo Simulations.
\newblock \emph{\bibinfo{journal}{Annual Review of Condensed Matter Physics}}
  \textbf{\bibinfo{volume}{4}}, \bibinfo{pages}{179--215}
  (\bibinfo{year}{2013}).

\bibitem{Xu16c}
\bibinfo{author}{Xu, X.~Y.}, \bibinfo{author}{Sun, K.},
  \bibinfo{author}{Schattner, Y.}, \bibinfo{author}{Berg, E.} \&
  \bibinfo{author}{Meng, Z.~Y.}
\newblock Non-Fermi Liquid at ($2+1$)$\mathrm{D}$ Ferromagnetic Quantum
  Critical Point.
\newblock \emph{\bibinfo{journal}{Phys. Rev. X}} \textbf{\bibinfo{volume}{7}},
  \bibinfo{pages}{031058} (\bibinfo{year}{2017}).

\bibitem{Assaad16}
\bibinfo{author}{Assaad, F.~F.} \& \bibinfo{author}{Grover, T.}
\newblock Simple Fermionic Model of Deconfined Phases and Phase Transitions.
\newblock \emph{\bibinfo{journal}{Phys. Rev. X}} \textbf{\bibinfo{volume}{6}},
  \bibinfo{pages}{041049} (\bibinfo{year}{2016}).

\bibitem{Gazit16}
\bibinfo{author}{Gazit, S.}, \bibinfo{author}{Randeria, M.} \&
  \bibinfo{author}{Vishwanath, A.}
\newblock Emergent Dirac fermions and broken symmetries in confined and
  deconfined phases of Z2 gauge theories.
\newblock \emph{\bibinfo{journal}{Nat. Phys.}} \textbf{\bibinfo{volume}{13}},
  \bibinfo{pages}{484--490} (\bibinfo{year}{2017}).

\bibitem{Gazit18}
\bibinfo{author}{Gazit, S.}, \bibinfo{author}{Assaad, F.~F.},
  \bibinfo{author}{Sachdev, S.}, \bibinfo{author}{Vishwanath, A.} \&
  \bibinfo{author}{Wang, C.}
\newblock Confinement transition of Z2 gauge theories coupled to massless
  fermions: Emergent quantum chromodynamics and SO(5) symmetry.
\newblock \emph{\bibinfo{journal}{Proceedings of the National Academy of
  Sciences}}  (\bibinfo{year}{2018}).

\bibitem{SatoT17}
\bibinfo{author}{Sato, T.}, \bibinfo{author}{Hohenadler, M.} \&
  \bibinfo{author}{Assaad, F.~F.}
\newblock Dirac Fermions with Competing Orders: Non-Landau Transition with
  Emergent Symmetry.
\newblock \emph{\bibinfo{journal}{Phys. Rev. Lett.}}
  \textbf{\bibinfo{volume}{119}}, \bibinfo{pages}{197203}
  (\bibinfo{year}{2017}).

\bibitem{Novoselov05}
\bibinfo{author}{Novoselov, K.~S.} \emph{et~al.}
\newblock Two-dimensional gas of massless Dirac fermions in graphene.
\newblock \emph{\bibinfo{journal}{Nature}} \textbf{\bibinfo{volume}{438}},
  \bibinfo{pages}{197--200} (\bibinfo{year}{2005}).

\bibitem{Blankenbecler81}
\bibinfo{author}{Blankenbecler, R.}, \bibinfo{author}{Scalapino, D.~J.} \&
  \bibinfo{author}{Sugar, R.~L.}
\newblock Monte Carlo calculations of coupled boson-fermion systems.
\newblock \emph{\bibinfo{journal}{Phys. Rev. D}} \textbf{\bibinfo{volume}{24}},
  \bibinfo{pages}{2278--2286} (\bibinfo{year}{1981}).

\bibitem{White89}
\bibinfo{author}{White, S.} \emph{et~al.}
\newblock Numerical study of the two-dimensional Hubbard model.
\newblock \emph{\bibinfo{journal}{Phys. Rev. B}} \textbf{\bibinfo{volume}{40}},
  \bibinfo{pages}{506--516} (\bibinfo{year}{1989}).

\bibitem{Assaad08_rev}
\bibinfo{author}{Assaad, F.} \& \bibinfo{author}{Evertz, H.}
\newblock World-line and Determinantal Quantum Monte Carlo Methods for Spins,
  Phonons and Electrons.
\newblock In \bibinfo{editor}{Fehske, H.}, \bibinfo{editor}{Schneider, R.} \&
  \bibinfo{editor}{Wei{\ss}e, A.} (eds.)
  \emph{\bibinfo{booktitle}{Computational Many-Particle Physics}}, vol.
  \bibinfo{volume}{739} of \emph{\bibinfo{series}{Lecture Notes in Physics}},
  \bibinfo{pages}{277--356} (\bibinfo{publisher}{Springer},
  \bibinfo{address}{Berlin Heidelberg}, \bibinfo{year}{2008}).

\bibitem{Capponi07}
\bibinfo{author}{Capponi, S.} \& \bibinfo{author}{Assaad, F.~F.}
\newblock Spin-nematic phases in models of correlated electron systems: A
  numerical study.
\newblock \emph{\bibinfo{journal}{Phys. Rev. B}} \textbf{\bibinfo{volume}{75}},
  \bibinfo{pages}{045115} (\bibinfo{year}{2007}).

\bibitem{Herbut09}
\bibinfo{author}{Herbut, I.~F.}, \bibinfo{author}{Juri\ifmmode \check{c}\else
  \v{c}\fi{}i\ifmmode~\acute{c}\else \'{c}\fi{}, V.} \& \bibinfo{author}{Roy,
  B.}
\newblock Theory of interacting electrons on the honeycomb lattice.
\newblock \emph{\bibinfo{journal}{Phys. Rev. B}} \textbf{\bibinfo{volume}{79}},
  \bibinfo{pages}{085116} (\bibinfo{year}{2009}).

\bibitem{Toldin14}
\bibinfo{author}{Parisen~Toldin, F.}, \bibinfo{author}{Hohenadler, M.},
  \bibinfo{author}{Assaad, F.~F.} \& \bibinfo{author}{Herbut, I.~F.}
\newblock Fermionic quantum criticality in honeycomb and $\pi$-flux Hubbard
  models: Finite-size scaling of renormalization-group-invariant observables
  from quantum Monte Carlo.
\newblock \emph{\bibinfo{journal}{Phys. Rev. B}} \textbf{\bibinfo{volume}{91}},
  \bibinfo{pages}{165108} (\bibinfo{year}{2015}).

\bibitem{Otsuka16}
\bibinfo{author}{Otsuka, Y.}, \bibinfo{author}{Yunoki, S.} \&
  \bibinfo{author}{Sorella, S.}
\newblock Universal Quantum Criticality in the Metal-Insulator Transition of
  Two-Dimensional Interacting Dirac Electrons.
\newblock \emph{\bibinfo{journal}{Phys. Rev. X}} \textbf{\bibinfo{volume}{6}},
  \bibinfo{pages}{011029} (\bibinfo{year}{2016}).

\bibitem{Zerf17}
\bibinfo{author}{Zerf, N.}, \bibinfo{author}{Mihaila, L.~N.},
  \bibinfo{author}{Marquard, P.}, \bibinfo{author}{Herbut, I.~F.} \&
  \bibinfo{author}{Scherer, M.~M.}
\newblock Four-loop critical exponents for the Gross-Neveu-Yukawa models.
\newblock \emph{\bibinfo{journal}{Phys. Rev. D}} \textbf{\bibinfo{volume}{96}},
  \bibinfo{pages}{096010} (\bibinfo{year}{2017}).

\bibitem{Assaad13}
\bibinfo{author}{Assaad, F.~F.} \& \bibinfo{author}{Herbut, I.~F.}
\newblock Pinning the Order: The Nature of Quantum Criticality in the Hubbard
  Model on Honeycomb Lattice.
\newblock \emph{\bibinfo{journal}{Phys. Rev. X}} \textbf{\bibinfo{volume}{3}},
  \bibinfo{pages}{031010} (\bibinfo{year}{2013}).

\bibitem{Ryu09}
\bibinfo{author}{Ryu, S.}, \bibinfo{author}{Mudry, C.}, \bibinfo{author}{Hou,
  C.-Y.} \& \bibinfo{author}{Chamon, C.}
\newblock Masses in graphenelike two-dimensional electronic systems:
  Topological defects in order parameters and their fractional exchange
  statistics.
\newblock \emph{\bibinfo{journal}{Phys. Rev. B}} \textbf{\bibinfo{volume}{80}},
  \bibinfo{pages}{205319} (\bibinfo{year}{2009}).

\bibitem{Abanov00}
\bibinfo{author}{Abanov, A.} \& \bibinfo{author}{Wiegmann, P.}
\newblock Theta-terms in nonlinear sigma-models.
\newblock \emph{\bibinfo{journal}{Nuclear Physics B}}
  \textbf{\bibinfo{volume}{570}}, \bibinfo{pages}{685 -- 698}
  (\bibinfo{year}{2000}).

\bibitem{Tanaka05}
\bibinfo{author}{Tanaka, A.} \& \bibinfo{author}{Hu, X.}
\newblock Many-Body Spin Berry Phases Emerging from the $\ensuremath{\pi}$-Flux
  State: Competition between Antiferromagnetism and the Valence-Bond-Solid
  State.
\newblock \emph{\bibinfo{journal}{Phys. Rev. Lett.}}
  \textbf{\bibinfo{volume}{95}}, \bibinfo{pages}{036402}
  (\bibinfo{year}{2005}).

\bibitem{Senthil06}
\bibinfo{author}{Senthil, T.} \& \bibinfo{author}{Fisher, M. P.~A.}
\newblock Competing orders, nonlinear sigma models, and topological terms in
  quantum magnets.
\newblock \emph{\bibinfo{journal}{Phys. Rev. B}} \textbf{\bibinfo{volume}{74}},
  \bibinfo{pages}{064405} (\bibinfo{year}{2006}).

\bibitem{Poland18}
\bibinfo{author}{{Poland}, D.}, \bibinfo{author}{{Rychkov}, S.} \&
  \bibinfo{author}{{Vichi}, A.}
\newblock {The Conformal Bootstrap: Numerical Techniques and Applications}.
\newblock \emph{\bibinfo{journal}{ArXiv:1805.04405}}  (\bibinfo{year}{2018}).

\bibitem{Nakayama16}
\bibinfo{author}{Nakayama, Y.} \& \bibinfo{author}{Ohtsuki, T.}
\newblock Necessary Condition for Emergent Symmetry from the Conformal
  Bootstrap.
\newblock \emph{\bibinfo{journal}{Phys. Rev. Lett.}}
  \textbf{\bibinfo{volume}{117}}, \bibinfo{pages}{131601}
  (\bibinfo{year}{2016}).

\bibitem{Kuklov08}
\bibinfo{author}{Kuklov, A.~B.}, \bibinfo{author}{Matsumoto, M.},
  \bibinfo{author}{Prokof'ev, N.~V.}, \bibinfo{author}{Svistunov, B.~V.} \&
  \bibinfo{author}{Troyer, M.}
\newblock Deconfined Criticality: Generic First-Order Transition in the SU(2)
  Symmetry Case.
\newblock \emph{\bibinfo{journal}{Phys. Rev. Lett.}}
  \textbf{\bibinfo{volume}{101}}, \bibinfo{pages}{050405}
  (\bibinfo{year}{2008}).

\bibitem{RychkovWalking1}
\bibinfo{author}{{Gorbenko}, V.}, \bibinfo{author}{{Rychkov}, S.} \&
  \bibinfo{author}{{Zan}, B.}
\newblock {Walking, Weak first-order transitions, and Complex CFTs}.
\newblock \emph{\bibinfo{journal}{ArXiv e-prints}}  (\bibinfo{year}{2018}).

\bibitem{Ippoliti18}
\bibinfo{author}{{Ippoliti}, M.}, \bibinfo{author}{{Mong}, R.~S.~K.},
  \bibinfo{author}{{Assaad}, F.~F.} \& \bibinfo{author}{{Zaletel}, M.~P.}
\newblock {Half-filled Landau levels: a continuum and sign-free regularization
  for 3D quantum critical points}.
\newblock \emph{\bibinfo{journal}{ArXiv:1810.00009}}  (\bibinfo{year}{2018}).

\bibitem{MetlitskiThorngren}
\bibinfo{author}{{Metlitski}, M.~A.} \& \bibinfo{author}{{Thorngren}, R.}
\newblock {Intrinsic and emergent anomalies at deconfined critical points}.
\newblock \emph{\bibinfo{journal}{\prb}} \textbf{\bibinfo{volume}{98}},
  \bibinfo{pages}{085140} (\bibinfo{year}{2018}).

\bibitem{Sandvik99b}
\bibinfo{author}{Sandvik, A.~W.}
\newblock Stochastic series expansion method with operator-loop update.
\newblock \emph{\bibinfo{journal}{Phys. Rev. B}} \textbf{\bibinfo{volume}{59}},
  \bibinfo{pages}{R14157--R14160} (\bibinfo{year}{1999}).

\bibitem{Sandvik11}
\bibinfo{author}{Sandvik, A.~W.}, \bibinfo{author}{Kotov, V.~N.} \&
  \bibinfo{author}{Sushkov, O.~P.}
\newblock Thermodynamics of a Gas of Deconfined Bosonic Spinons in Two
  Dimensions.
\newblock \emph{\bibinfo{journal}{Phys. Rev. Lett.}}
  \textbf{\bibinfo{volume}{106}}, \bibinfo{pages}{207203}
  (\bibinfo{year}{2011}).

\bibitem{Janssen18}
\bibinfo{author}{Janssen, L.}, \bibinfo{author}{Herbut, I.~F.} \&
  \bibinfo{author}{Scherer, M.~M.}
\newblock Compatible orders and fermion-induced emergent symmetry in Dirac
  systems.
\newblock \emph{\bibinfo{journal}{Phys. Rev. B}} \textbf{\bibinfo{volume}{97}},
  \bibinfo{pages}{041117} (\bibinfo{year}{2018}).

\bibitem{ALF_v1}
\bibinfo{author}{Bercx, M.}, \bibinfo{author}{Goth, F.},
  \bibinfo{author}{Hofmann, J.~S.} \& \bibinfo{author}{Assaad, F.~F.}
\newblock {The ALF (Algorithms for Lattice Fermions) project release 1.0.
  Documentation for the auxiliary field quantum Monte Carlo code}.
\newblock \emph{\bibinfo{journal}{SciPost Phys.}} \textbf{\bibinfo{volume}{3}},
  \bibinfo{pages}{013} (\bibinfo{year}{2017}).

\bibitem{Wu04}
\bibinfo{author}{Wu, C.} \& \bibinfo{author}{Zhang, S.-C.}
\newblock Sufficient condition for absence of the sign problem in the fermionic
  quantum Monte Carlo algorithm.
\newblock \emph{\bibinfo{journal}{Phys. Rev. B}} \textbf{\bibinfo{volume}{71}},
  \bibinfo{pages}{155115} (\bibinfo{year}{2005}).

\bibitem{Li16}
\bibinfo{author}{Li, Z.-X.}, \bibinfo{author}{Jiang, Y.-F.} \&
  \bibinfo{author}{Yao, H.}
\newblock Majorana-Time-Reversal Symmetries: A Fundamental Principle for
  Sign-Problem-Free Quantum Monte Carlo Simulations.
\newblock \emph{\bibinfo{journal}{Phys. Rev. Lett.}}
  \textbf{\bibinfo{volume}{117}}, \bibinfo{pages}{267002}
  (\bibinfo{year}{2016}).

\bibitem{Wei16}
\bibinfo{author}{Wei, Z.~C.}, \bibinfo{author}{Wu, C.}, \bibinfo{author}{Li,
  Y.}, \bibinfo{author}{Zhang, S.} \& \bibinfo{author}{Xiang, T.}
\newblock Majorana Positivity and the Fermion Sign Problem of Quantum Monte
  Carlo Simulations.
\newblock \emph{\bibinfo{journal}{Phys. Rev. Lett.}}
  \textbf{\bibinfo{volume}{116}}, \bibinfo{pages}{250601}
  (\bibinfo{year}{2016}).

\bibitem{Hofmann18}
\bibinfo{author}{{Hofmann}, J.~S.}, \bibinfo{author}{{Assaad}, F.~F.} \&
  \bibinfo{author}{{Grover}, T.}
\newblock {Kondo Breakdown via Fractionalization in a Frustrated Kondo Lattice
  Model}.
\newblock \emph{\bibinfo{journal}{ArXiv:1807.08202}}  (\bibinfo{year}{2018}).

\end{thebibliography}
\end{document}